\documentclass[draft]{agujournal2019}
\usepackage{url} 
\usepackage{soul}
\draftfalse

\journalname{JGR: Atmospheres}

\begin{document}

%
%


\title{Effective Emission Heights of Various OH Lines From X-shooter and SABER 
Observations of a Passing Quasi-2-Day Wave}

%
%




\authors{S. Noll\affil{1,2}, C. Schmidt\affil{2}, W. Kausch\affil{3}, 
  M. Bittner\affil{2,1}, S. Kimeswenger\affil{3,4}}


\affiliation{1}{Institut f\"ur Physik, Universit\"at Augsburg, Augsburg,
  Germany}
\affiliation{2}{Deutsches Fernerkundungsdatenzentrum, Deutsches Zentrum f\"ur 
  Luft- und Raumfahrt, We\ss{}ling-Oberpfaffenhofen, Germany}
\affiliation{3}{Institut f\"ur Astro- und Teilchenphysik, Universit\"at
  Innsbruck, Innsbruck, Austria}
\affiliation{4}{Instituto de Astronom\'ia, Universidad Cat\'olica del Norte,
  Antofagasta, Chile}





\correspondingauthor{Stefan Noll}{stefan.noll@dlr.de}




\begin{keypoints}
\item X-shooter-based intensities of 298 OH lines from eight nights show a 
strong Q2DW in southern summer 2017 
\item Fits of the Q2DW phase in the X-shooter data and SABER-based OH 
emission profiles were used to derive effective OH emission heights
\item The line-dependent wave amplitudes confirm the presence of cold and
hot OH populations for each vibrational level 
\end{keypoints}

%
%

%
%


\begin{abstract}
Chemiluminescent radiation of the vibrationally and rotationally excited OH
radical, which dominates the nighttime near-infrared emission of the Earth's
atmosphere in wide wavelength regions, is an important tracer of the chemical
and dynamical state of the mesopause region between 80 and 100\,km. As 
radiative lifetimes and rate coefficients for collision-related transitions
depend on the OH energy level, line-dependent emission profiles are expected. 
However, except for some height differences for whole bands mostly revealed 
by satellite-based measurements, there is a lack of data for individual 
lines. We succeeded in deriving effective emission heights for 298 OH lines 
thanks to the joint observation of a strong quasi-2-day wave (Q2DW) in eight 
nights in 2017 with the medium-resolution spectrograph X-shooter at the Very 
Large Telescope at Cerro Paranal in Chile and the limb-sounding SABER 
radiometer on the TIMED satellite. Our fitting procedure revealed the most
convincing results for a single wave with a period of about 44\,h and a 
vertical wavelength of about 32\,km. The line-dependent as well as 
altitude-resolved phases of the Q2DW then resulted in effective heights which 
differ by up to 8\,km and tend to increase with increasing vibrational and 
rotational excitation. The measured dependence of emission heights and wave 
amplitudes (which were strongest after midnight) on the line parameters 
implies the presence of a cold thermalized and a hot non-thermalized 
population for each vibrational level. 
\end{abstract}

\section*{Plain Language Summary}
Hydroxyl (OH) is an important molecule in the Earth's atmosphere at altitudes
between 80 and 100\,km. It is the main source of atmospheric nighttime 
radiation in the near-infrared wavelength range and is therefore a valuable
tracer of the chemistry and dynamics at high altitudes. The emission spectrum
consists of various lines which are related to different levels of vibration 
and rotation. Although the vertical emission distribution should depend on 
the given line due to differences in the deactivation of the corresponding
energy levels, the line-specific details have been uncertain until now. We 
have succeeded in deriving effective time-averaged emission heights for 298 
OH lines based on the combination of ground-based line-resolved and
space-based height-resolved observations of a very strong rising wave with a 
period close to 2 days and a relatively short vertical wavelength in eight 
nights in 2017. The resulting heights (obtained via the line-dependent wave 
phases) differ by up to 8 km and generally increase with higher molecular 
vibration and rotation. They are valuable for ground-based studies of other 
waves and contribute (combined with conclusions from the wave amplitudes) to 
a better understanding of the internal processes in OH molecules.

%
%

%


%
%
%
%

\section{Introduction}\label{sec:intro}

The nocturnal atmospheric emission of wide wavelength regions in the 
near-infrared is dominated by chemiluminescent radiation of various 
roto-vibrational bands up to the vibrational level $v = 9$ of the electronic 
ground state of the hydroxyl (OH) radical 
\cite<e.g.,>{meinel50,noll15,rousselot00}. Excited OH is mostly produced by 
the reaction of hydrogen with ozone \cite{bates50}. The nightglow emission 
essentially originates from altitudes between 80 and 100\,km, and is 
therefore an important tracer of the chemistry and dynamics in the Earth's 
mesopause region. Rocket flights showed typical peak heights of about 87\,km 
and layer widths of about 8\,km \cite{baker88}. Moreover, OH emission 
profiles have frequently been observed by limb-sounding instruments in space 
\cite<e.g.,>{baker07,dodd94,savigny12,wuest20,yee97}. Rare ground-based 
altitude measurements relied on the identification of the same emission 
features in imaging instruments at different sites 
\cite<e.g.,>{kubota99,moreels08}. \citeA{yu17} combined wind observations 
with a meteor radar and a Fabry-Perot interferometer focusing on OH to 
estimate peak heights of the radiation. Finally, rough estimates are also 
possible by means of the well-established negative relation between effective 
emission height and emission intensity 
\cite<e.g.,>{garcia17,liu06,mulligan09,savigny15,yee97}, which can be 
explained by the larger change of the layer profile at lower altitudes due to 
the stronger relative variations of atomic oxygen (which is required for the 
ozone production). The published relations had to be calibrated with 
satellite observations.  

The hydrogen--ozone reaction mainly produces OH populations in high $v$ 
between 7 and 9 \cite<e.g.,>{adler97}. In contrast, measured populations 
increase with decreasing $v$ \cite<e.g.,>{takahashi81,cosby07,noll15}. This
discrepancy is caused by the step-by-step relaxation process due to the 
radiation of photons and collisions with other atmospheric constituents 
\cite<e.g.,>{adler97,xu12,noll18b}. As the radiative lifetimes and rate
coefficients for the collisions with different species depend on the OH 
level, the resulting emission profiles for individual lines should differ.  
Models show an increase of the peak emission height with increasing $v$
\cite<e.g.,>{adler97,makhlouf95,mcdade91,savigny12,xu12}. The spread between
high and low $v$ may amount to several kilometers with possibly larger 
height shifts for low $v$. Collisions of OH with atomic oxygen appear to
be especially crucial for the $v$-dependent discrepancies \cite{savigny12}. 
The impact of rotational energy on the altitude distribution has been 
modeled quite rarely. \citeA{dodd94} discussed the differences between 
populations with rotational quantum numbers $N$ up to 7 and those with 
$N \ge 11$. Density profiles for the larger $N$ appear to peak higher with 
similar altitude differences as obtained for the comparison of low and high 
$v$. Moreover, the density peaks seem to show less variation for larger $N$
with respect to changes in $v$. \citeA{noll18b} presented modeling results
for $v = 9$ and found maximum $N$-dependent height differences between 1.5
and 2.8\,km depending on the uncertain rate coefficients for collisions with
atomic oxygen. 

Measurements of the impact of the OH energy state on the height distribution 
were successfully performed with space-based limb sounding. The studies 
mostly focused on the comparison of the emission profiles of a few 
roto-vibrational bands \cite{noll16,sheese14,savigny13,savigny12}. The 
results suggest typical height changes for $\Delta v = 1$ of about 0.4 to 
0.5\,km. These differences can vary by a significant fraction of the values. 
They tend to increase with lower peak altitude and higher atomic oxygen 
concentration \cite{savigny13}. The former can change by several kilometers 
forced by waves on different time scales and the general circulation 
\cite<e.g.,>{liu06,nikoukar07,noll16,teiser17,yee97}. The majority of the 
variability is found below the emission peak \cite{nikoukar07}, where the
atomic oxygen density profile is particularly steep. In contrast to whole
roto-vibrational bands, there is a lack of studies of individual lines with
different $N$. The only noteworthy data known to us are related to 
observations with the Cryogenic InfraRed Radiance Instrumentation for Shuttle 
(CIRRIS 1A) Michelson interferometer onboard Space Shuttle Discovery for 
three days in 1991 \cite{dodd94}. The measured limb profiles comprise lines 
of several $v$ with $N \le 4$ as well as purely rotational lines with $N$ 
around 15 and 30. However, the interpretation of the data required 
sophisticated modeling.  
 
There is an alternative approach to estimate effective emission heights of
OH lines. Perturbations passing the mesopause region will produce 
characteristic patterns in the OH intensity time series that will be shifted
in time depending on the altitude of the emission \cite{noll15,schmidt18}. In 
this way, the relative layering of different emissions can be obtained from 
ground-based observations of individual lines. Nevertheless, the derivation 
of absolute heights will also need information on the vertical propagation of 
the perturbation through the OH emission layer as can be measured by 
satellite-based limb-sounding instruments. As the satellites relevant for 
nightglow observations have nearly Sun-synchronous orbits 
\cite<e.g.,>{russell99,savigny12,yee97}, the minimum time scale of the 
variations needs to be of the order of days for a promising combination of
the profile data with ground-based spectra of OH lines in order to link wave 
phases with altitudes in the same region.   

The so-called quasi-2-day wave (Q2DW) can achieve very high amplitudes in the
mesopause region at low to middle latitudes. In particular, strong Q2DWs 
occur in the Southern Hemisphere for several weeks in the summer months 
January and February \cite<e.g.,>{ern13,gu19,tunbridge11,walterscheid15}. The 
period is close to 2 days but can vary from 42 to 54\,h \cite{gu19}. For 
periods very close to 48\,h, there can be phase locking with tidal modes
\cite{walterscheid96}. Southern Q2DWs usually show a dominating 
westward-moving longitudinal pattern with a zonal wavenumber of 3 (W3), which
can be accompanied by other modes with different periods, zonal wavenumbers, 
and propagation directions \cite<e.g.,>{he21,pedatella12,tunbridge11}. Q2DWs 
are regarded to belong to the class of Rossby-gravity waves 
\cite<e.g.,>{salby81}. Their genesis is probably related to 
baroclinic/barotropic instabilities in the summertime easterly zonal wind jet 
in the lower mesosphere \cite{plumb83}, which seem to be linked to increased 
gravity-wave drag \cite{ern13}. The amplitudes maximize at the OH emission 
heights and can reach more than 10\,K in temperature \cite{gu19,tunbridge11} 
and 50\,m\,s$^{-1}$ in wind \cite{limpasuvan05,wu93}. Between the Q2DW source 
region and these altitudes at low to middle latitudes, the propagation 
perpendicular to the zonal component is effectively upward. The situation is 
different in the lower thermosphere well above the OH layer, where a 
northward meridional flow tends to be dominant in the Southern Hemisphere 
\cite{yue12}. At OH emission heights, vertical wavelengths are mostly longer 
than 20\,km and are often even beyond 100\,km \cite{huang13,reisin21}. The 
impact of a Q2DW on OH emission was investigated by \citeA{pedatella12} based 
on space-based OH profile data for January 2006. The peak emission rate 
varied by a factor of about 4. This large effect appears to be mostly related 
to Q2DW-induced variations of the atomic oxygen concentration at the OH 
emission heights. The concentration can also be modulated by tides and 
gravity waves. 

We have succeeded in deriving effective emission heights for about 300 OH 
lines by means of the combination of ground-based spectroscopic and 
space-based height-resolved observations of a strong Q2DW in January to
February 2017. The OH line intensities originate from near-infrared spectra of 
the medium-resolution spectrograph X-shooter \cite{vernet11} of the Very 
Large Telescope (VLT) of the European Southern Observatory (ESO) at Cerro 
Paranal in Chile (24.6$^{\circ}$\,S, 70.4$^{\circ}$\,W). Scans of the OH 
emission profiles in two filter bands at about 1.6 and 2.1\,$\mu$m are related
to measurements with the Sounding of the Atmosphere using Broadband Emission 
Radiometry (SABER) instrument onboard the Thermosphere Ionosphere Mesosphere 
Energetics Dynamics (TIMED) satellite \cite{russell99}. In the following 
section~\ref{sec:data}, we will describe the data sets for both instruments. 
Then, we will present our approach to fit the observed wave 
(section~\ref{sec:methods}). Our results will be shown in 
section~\ref{sec:results}, which also includes our estimates of line-dependent
effective emission heights (section~\ref{sec:heights}) and a brief discussion 
of another Q2DW in 2019 (section~\ref{sec:2019}). Finally, we will draw our 
conclusions (section~\ref{sec:conclusions}).

\section{Data}\label{sec:data}

\subsection{X-shooter}\label{sec:xshooter}

The X-shooter medium-resolution echelle spectrograph \cite{vernet11}
mounted at an 8\,m telescope of the VLT can simultaneously observe a very 
wide wavelength range from 0.3 to 2.5\,$\mu$m covered by three spectroscopic 
arms with separate optical components and detectors. Since the strongest OH 
bands for all upper vibrational levels $v^{\prime}$ from 2 to 9 are found in 
the near-infrared regime \cite<e.g.,>{rousselot00}, we focus our analysis on 
the correspondingly named NIR arm, which extends from 1.0 to 
2.5\,$\mu$m. For this arm and scientific targets, the projected slit width 
can vary between 0.4 and 1.5$^{\prime\prime}$ on the sky, which corresponds 
to a spectral resolving power between 12,000 and 3,500. The projected slit 
length is fixed to 11$^{\prime\prime}$.

The spectra used originate from the ESO Science Archive Facility. This study 
only uses a small fraction of the data that were processed by us. The whole 
sample comprises about 90,000 spectra in the NIR arm with an exposure time of 
at least 10\,s that were taken between October 2009 and September 2019. The 
basic processing of the raw echelle spectra was performed by means of the 
official reduction pipeline \cite{modigliani10} in version v2.6.8 and 
calibration data preprocessed by ESO \cite<see also>{unterguggenberger17}. 
The resulting two-dimensional (2D) wavelength-calibrated sky spectra were 
subjected to post-processing optimized for the retrieval of airglow emission. 
First, the 2D spectra were averaged along the spatial direction in order to 
obtain one-dimensional (1D) spectra. Then, the separation of sky emission and 
astronomical target performed by the pipeline was improved. For this goal, we 
extracted residual sky in the 2D object spectrum from the third of the slit 
positions with the lowest integrated counts in the whole wavelength range. 
After the minimization of noise by smoothing, we added the resulting 1D
mean spectrum to the sky spectrum. This approach was particularly useful for 
bright star-like astronomical targets.

Although the pipeline produces flux-calibrated spectra, we performed this 
processing step independently in order to minimize the errors by means of a 
consistent approach for the entire data set. For the NIR arm, 
pipeline-processed spectra of the relatively bright spectrophotometric 
standard stars EG\,274 and LTT\,3218 \cite{moehler14} were corrected for 
telluric absorption using a model-based fitting approach 
\cite{smette15,kausch15} and then compared to the theoretical spectral energy 
distributions in order to derive individual response curves that indicate the 
wavelength-dependent instrumental quantum efficiency. As the resulting 
response curves for both stars did not exactly match, we lowered the quantum 
efficiency of the EG\,274 curves (by about 3\% in a wide wavelength range) 
for a better consistency. Thus, remaining systematic uncertainties in the 
absolute flux calibration for clear sky conditions mainly depend on the 
quality of the reference spectrum for LTT\,3218. As the response curves are 
relatively uncertain at wavelengths longer than 1.9\,$\mu$m 
\cite<see>{noll15}, we used spectra of so-called telluric standard stars with 
well-known spectral energy distributions (Rayleigh--Jeans law) for 
atmospheric absorption measurements to derive a general correction function. 
In order to further improve the quality, master response curves for 10 time 
periods with lengths between 9 and 15 months were created including only the 
best-quality data and involving a final smoothing procedure. The splitting of 
the periods is usually at New Year but also considers a change in the 
calibration products in January 2013 and a recoating of the main mirror in 
December 2016. An analysis of the variability of the flux-calibrated star 
spectra points to relative uncertainties of only 2 to 3\% up to 2.1\,$\mu$m 
in the NIR arm. This is a clear improvement in comparison to a maximum 
difference in the response of about 12\%. The final set of master response 
curves was applied to the 1D sky spectra. A minor fraction of the spectra 
were taken with a so-called $K$-blocking filter \cite{vernet11}, which 
reduces the useful wavelength range to wavelengths shorter than 2.1\,$\mu$m. 
These spectra had to be calibrated with a 1.3 times higher response in order 
to correctly consider differences in the calibration data.        

The measurement of the line intensities was performed in two steps. First,
the underlying continuum was determined. After various tests, our procedure
used the first quintile of the pixel intensities (cut at 20\%) in wavelength
ranges with a relative width of 0.008 to derive the continuum at the 
corresponding central positions. The width was increased by the factors 5.0 
and 2.5 around the dense roto-vibrational O$_2$ emission bands at about 1.27 
and 1.58\,$\mu$m, respectively \cite<e.g.,>{rousselot00}. It was multiplied 
by a factor of 0.4 beyond 2.08\,$\mu$m in order to avoid issues with the 
steeply increasing continuum due to thermal radiation of the telescope. After
subtraction of the resulting continuum, the residual flux was integrated in 
specific wavelength ranges depending on the central wavelength of the OH
$\Lambda$ doublet, separation of both components \cite<e.g.,>{noll20}, and 
slit width. As an X-shooter profile of an unresolved line doublet (which is 
true in most cases) can be approximated by the combination of a fixed Gaussian 
and a slit-dependent boxcar, the integration limits had been optimized to 
assure relatively stable distances to the positions marking half peak 
intensity. For example, the size of these margins amounts to about 60\% of the 
full integration range minus the $\Lambda$ doublet separation for the most 
frequently used slit with a width of 0.9$^{\prime\prime}$. The separation of
the integration limits was sufficiently wide to avoid significant flux losses
by uncertainties in the wavelength calibration (fraction of a pixel). 
Positions of suitable lines were taken from \citeA{brooke16}. Important 
selection criteria were clear detectability (at least for long exposures), 
the lack of blends with other emission lines, a smooth continuum without 
strong absorption features, and high atmospheric transmission. In the end,
the selection procedure (which also included a check for outliers in the 
scientific analysis) resulted in 298 OH $\Lambda$ doublets from 14 bands, 
which cover upper vibrational levels $v^{\prime}$ between 2 and 9 and upper 
rotational levels $N^{\prime}$ up to 15.

The measured line intensities were corrected for the van Rhijn effect
\cite{vanrhijn21}, i.e. the projected layer width, assuming a reference 
altitude of 87\,km. The correction of this effect is important as there is
a strong variation in the zenith angles of astronomical observations. We also
considered the absorption of line emission by molecules in the lower 
atmosphere in a similar way as decribed by \citeA{noll15}. The approach 
involves the calculation of line-specific transmissions based on the zenith
angle, Cerro Paranal atmospheric data \cite{noll12}, the Line-By-Line 
Radiative Transfer Model \cite<LBLRTM,>{clough05}, and the assumption of 
purely Doppler-broadened OH lines for a temperature of 190\,K at 
\citeA{brooke16} wavelengths. Transmissions for entire $\Lambda$ doublets 
were derived by means of weights for the individual components that were 
taken from the branch-specific OH level population fits of \citeA{noll20}. 
Reference line transmissions were calculated for zenith and a typical amount 
of precipitable water vapour (PWV) of 2.5\,mm. Doublets with reference 
transmissions lower than 70\% were not considered for the final line set
(average of 96\%). Differences in the optical depth in the spectroscopic data 
set were calculated depending on the most crucial parameters, zenith angle and 
PWV. For the latter, accurate measurements with a Low Humidity And Temperature 
PROfiler (L-HATPRO) microwave radiometer are available at Cerro Paranal 
\cite{kerber12}. However, as the L-HATPRO data set starts in 2014 and shows
several gaps afterwards, we mostly used these data to calibrate intensity 
ratios of lines with low and high transmission in the bands OH(2-0) and 
OH(6-4) to estimate PWV values for each spectrum \cite<cf.>{xu20}. Up to
very high L-HATPRO PWVs of about 16\,mm, the approach works quite well with a 
mean offset of -0.1\,mm and a standard deviation of 0.6\,mm. With an upper 
cut at 6\,mm (which still represents about 90\% of the data), the systematic 
shift is negligible and the scatter is only half as large.

The quality of the measurement of OH line intensities strongly varies from
spectrum to spectrum due to strong changes in spectral resolution, exposure 
time, atmospheric water vapour, and especially contamination by 
(spatially extended) astronomical targets that can have a different effect on 
each line. For this reason, we performed a line-specific selection of 
suitable observations. Starting with a preselected set of 88,481 useful 
spectra, we carried out an iterative $\sigma$-clipping procedure for each 
line. This outlier rejection involved limits with respect to the underlying 
continuum, the intensity error, and the intensity itself. The error was 
estimated from the deviation of the residual continuum at both margins of the 
integration range from the zero line. It was made sure that it was always 
lower than 10\% for the selected measurements. As a result, the line-specific 
selection rates varied between 69.4\% and 99.8\% with a mean value of 94.1\% 
for wavelengths shorter than 2.1\,$\mu$m. In order to handle the wide range 
of exposure times between 10\,s and 2.5\,h and the corresponding impact on 
the data quality, we did not use the individual measurements for the 
analysis. Instead, we divided the 10 years into bins of 30\,min, which are 
sufficiently short for the analysis of Q2DWs. Each bin contains the 
line-specific spectra with matching central time. The effective intensities 
were calculated weighted by the exposure time. In the subsequent analysis, we 
only considered those bins with a summed exposure time of at least 10\,min. 
The average numbers of the selected bins were 19,480 and 17,001 for lines at 
wavelengths shorter and longer than 2.1\,$\mu$m. A lower minimum exposure 
time of 5\,min did not significantly change the results of the analysis but 
increased the scatter. 

The resulting binned time series were inspected with respect to Q2DW 
features. The focus was on the months January and February. The 
X-shooter data set shows a strongly varying time coverage due to the
different astronomical observing programs and the usual mounting of two
to three instruments at the same telescope, which can cause gaps of the order 
of weeks. As a consequence, we clearly identified Q2DW events only in the
years 2017 and 2019. Moreover, only fractions of the lifetime of a wave were
well covered by X-shooter observations. The good intervals are eight 
nights from 26 January to 3 February 2017 and seven nights from 11 to 18
January 2019. The corresponding maximum numbers of bins of 30\,min are 88
and 89, respectively. As the sample of selected spectra is different for each 
line, up to three bins from 2017 were lost for line wavelengths below 
2.1\,$\mu$m. At longer wavelengths, where 30 OH lines are affected by the 
optional $K$-blocking filter, the bin numbers were 78 and 35 for the years 
2017 and 2019, respectively. The comparison of Q2DW results of lines with the 
same upper levels but different bin numbers did not show any significant 
discrepancies for the data from 2017. However, the lines at long wavelengths
had to be excluded from the analysis of the data from 2019. The drop from 89 
to 35 bins was too large. 

Our discussion of the results in section~\ref{sec:results} will focus on the
event in 2017 since only these data were suitable to estimate effective 
emission heights of OH lines, i.e. the main goal of this study. The issues
related to the data from 2019 will briefly be described in 
section~\ref{sec:2019}.

\subsection{SABER}\label{sec:saber}

For better understanding the Q2DW-related X-shooter data and adding
important altitude-dependent information, we also considered limb-sounding
data (version v2.0) of the SABER radiometer on TIMED \cite{russell99}. The
observing channels centered on 1.64 and 2.06\,$\mu$m essentially cover 
emission of the OH(4-2) and OH(5-3) bands as well as the OH(8-6) and OH(9-7) 
bands \cite{baker07}. Consequently, the effective upper vibrational levels of
both channels are about 4.6 and 8.3 \cite{noll16}. We used the so-called
``unfiltered'' products, i.e. the given volume emission rates (VERs) had been
corrected for the missing line emission of the stated OH bands in the 
wavelength ranges covered by the channels \cite{mlynczak05}. For some checks, 
we also considered VER profiles of the channel at about 1.27\,$\mu$m, which 
mostly comprises the O$_2$(a-X)(0-0) emission band 
\cite<e.g.,>{noll16,rousselot00}. Moreover, we used the kinetic temperature 
products, which are based on CO$_2$ observations at 15\,$\mu$m combined with 
modeling \cite{dawkins18,remsberg08}. All profiles were interpolated to match 
the same vertical grid with a step size of 0.2\,km. The natural vertical 
resolution of about 2\,km for the relevant altitude range smoothes the 
profiles, but estimates of the peak altitude are possible with much higher 
accuracy. For the two OH channels, we also calculated vertically integrated 
VERs, which should correlate with the summed X-shooter line intensities in 
the same wavelength range. The integration was limited to altitudes that 
did not deviate more than 15\,km from the mean of the two heights with half 
peak emission, which was often several hundred meters higher than the 
emission peak.

As the X-shooter data just revealed clear Q2DW events in 2017 and 
2019 (section~\ref{sec:xshooter}), we only selected the archived SABER data 
from January and February of both years. The spatial selection was performed 
for a latitude band centered on Cerro Paranal with a full width of 
10$^{\circ}$. In terms of the longitude, the width was 20$^{\circ}$. As the 
satellite moves during single scans, the reference selection coordinates were 
taken at 87\,km. The size of the geographical area is a compromise between 
minimization of spatial effects and maximization of the sample size 
\cite<see>[and the discussion in section~\ref{sec:profiles}]{noll16}. For the 
period from 26 January to 3 February 2017, the described limits resulted in 
44 profiles. As the nearly Sun-synchronous orbit of TIMED precesses with a 
period of about 60 days \cite{russell99}, the available coverage of local 
time (LT) is very limited for our study. In the case of the eight nights from 
2017, the coverage is restricted to LTs of about 21:00 and about 04:00. 
Each time window includes half of the profiles. For the seven nights from 11 
to 18 January 2019, only 19 profiles with LTs close to 23:00 from the 
descending node of the TIMED orbit could be used. Measurements at about 06:00 
could not be taken as the Sun was above the horizon. The latter significantly 
changes the OH emission with respect to total intensity and effective layer 
height \cite<e.g.,>{yee97}.

\section{Methods}\label{sec:methods}

For the derivation of effective emission heights, we need to fit the Q2DW in
the time series of X-shooter-based OH line intensities 
(section~\ref{sec:xshooter}) and broad-band SABER emission rates 
(section~\ref{sec:saber}) with a suitable wave model. In the end, it is 
important that the resulting effective wave phases for the various OH lines 
can be linked to altitudes via a SABER-based monotonic phase--height relation 
with a significant range of phases for the same wave period. Moreover, the
model needs to be sufficiently robust to produce consistent fits for all OH
lines and all relevant altitude levels. As long as these conditions are 
fulfilled, it is not essential that the model is particularly accurate with 
respect to the real wave properties, which could be quite complex. Even a 
model with very different wave parameters might work (see 
section~\ref{sec:heights}). We therefore applied a simple periodic cosine 
function. However, we did not use the entire data set for the fits as it 
turned out that especially the important data from 2017 showed a strong 
dependence of the wave amplitude on local time, which is probably related to 
significant interactions between the Q2DW and tides (see 
section~\ref{sec:lines}). In order to avoid a line-dependent systematic bias, 
we fitted only data points with similar LT and combined the reliable fits 
with respect to the wave phases afterwards. The fitting of the LT dependence 
of the amplitude by a second wave did not work as the variability pattern 
could not be reproduced by a simple trigonometric function. Motivated by 
previous results on Q2DWs (see section~\ref{sec:intro}), we also tested a 
model with two Q2DW components using the whole data set. This model was only 
partly able to fit the time series. Moreover, the best wave parameters were 
not convincing as very similar amplitudes and vertical wavelengths were 
needed for components with different periods (above and below 48\,h) and 
opposite vertical propagation directions. The introduction of additional wave 
components could improve the fits. However, the increased number of fit 
parameters could negatively affect the robustness of the fits and the 
derivation of useful phases for the emission height estimates. 

For our final wave fits, we used the fit formula
\begin{linenomath*}
\begin{equation}\label{eq:fit}
f(t, t_\mathrm{LT}) = 
c(t_\mathrm{LT}) \left(a(t_\mathrm{LT}) \cos\left(2\pi 
\left(\frac{t}{T}\,–\,k\,\Delta\lambda\,-\,\phi\right)\right) + 1\right),
\end{equation}
\end{linenomath*}
where $t$ is the time of the observation relative to a reference. For the 
Q2DW in 2017, we used 30 January 2017 12:00 LT (mean solar noon) at Cerro 
Paranal, which is in the middle of the time interval. For 2019, the reference 
was 15 January 2019 12:00 LT. The period of the wave is $T$ and a relative 
amplitude in the range from 0 to 1 is given by $a$. The latter depends on the 
LT-related data selection, which is marked by $t_\mathrm{LT}$. This also 
applies to the factor $c$, which can alternatively be interpreted as an 
additive constant. As we divided all data by the average of the respective 
times series before the fit, the product $c{\cdot}a$ represents the wave 
amplitude relative to the sample mean. Moreover, $c$ values clearly different 
from 1 indicate additional variability beyond the Q2DW and/or limitations in 
the assumption of a cosine function. The latter is not unlikely in the view 
of the huge amplitudes found (see section~\ref{sec:results}). The 
height-dependent phase $\phi$ is defined relative to $T$, i.e. it is a 
fraction without unit. As the temporal term in the formula is positive, the
minus sign in front of $\phi$ is needed for the correct definition
\cite<cf.>{forbes95}. The phase is given for the longitude of Cerro Paranal 
$\lambda_\mathrm{CP}$ relative to the full circle. Deviations $\Delta\lambda$
from $\lambda_\mathrm{CP}$ cause an influence of the zonal wavenumber $k$ on 
the fit. Westward wave propagation is marked by negative $k$. As the 
dominating Q2DW mode in the southern hemisphere is W3 (see 
section~\ref{sec:intro}), we used $k = -3$. This choice was confirmed by 
additional fits with free $k$ of a SABER sample without longitude and LT 
restriction, which also agreed with the SABER-based results of \citeA{gu19} 
for 2017. For the local X-shooter data, we always set $\Delta\lambda = 0$. 
For the fits of the satellite-based SABER data, the zonal term is a small but 
significant correction with average positive and negative $\Delta\lambda$ of 
about 0.014 (i.e. 5$^{\circ}$) for 2017 and about 0.012 for 2019. The fitting 
was performed with a least-squares algorithm involving bounds (e.g. for 
excluding negative amplitudes). After some tests with $T$ as a free fit 
parameter, we decided to make separate fits for a grid of periods $T$ from 40
to 60\,h with a step size of 1\,h. This approach significantly improved the 
robustness of the fits and made sure that the resulting wave phases for the 
different OH emissions referred to the same $T$. Consequently, the only free 
parameters for each fit were $c$, $a$, and $\phi$. 

\begin{table}
\caption{Derivation of optimum Q2DW phase relative to a period of 44\,h at
30 January 2017 12:00 LT at Cerro Paranal for X-shooter-based relative 
OH(4-2)P$_1$(1) intensities} 
\centering
\begin{tabular}{c c c c c c}
\hline
LT bin & Sample & Phase & Error$^{a}$ & Weight$^{b}$ \\
\hline
$20 - 21$ &  6 & 0.733 & 0.131 & 0.000 \\
$21 - 22$ & 11 & 0.847 & 0.554 & 0.000 \\
$22 - 23$ &  8 & 0.426 & 0.369 & 0.000 \\
$23 - 24$ & 11 & 0.406 & 0.104 & 0.000 \\
$00 - 01$ & 12 & 0.406 & 0.049 & 0.127 \\
$01 - 02$ & 13 & 0.381 & 0.031 & 0.201 \\
$02 - 03$ & 14 & 0.388 & 0.022 & 0.285 \\
$03 - 04$ & 12 & 0.423 & 0.016 & 0.387 \\
$04 - 05$ &  1 & 0.000 & 9.999 & 0.000 \\
\hline
Average   & 51 & 0.403 & 0.018 & 1.000 \\
\hline
\multicolumn{5}{l}{$^{a}$Fit uncertainty for LT bins and weighted} \\ 
\multicolumn{5}{l}{standard deviation of hour-specific phases in} \\
\multicolumn{5}{l}{the case of the weighted average.} \\
\multicolumn{5}{l}{$^{b}$Weight of zero for LT bins with less than} \\
\multicolumn{5}{l}{10 data points and/or a phase error more} \\
\multicolumn{5}{l}{than 5 times higher than the minimum.}
\end{tabular}
\label{tab:xs_phase}
\end{table}

For the X-shooter data consisting of relative intensities of 30\,min bins for 
each line, the fitting procedure included three steps. First, we used all 
data of each line for a rough initial estimate of the phase $\phi$. In the 
next step, we performed separate fits for each hour of the night with at 
least five available bins. The hour-dependent fits were then used to derive 
an optimum phase for all data. Table~\ref{tab:xs_phase} illustrates this 
procedure for OH(4-2)P$_1$($N^{\prime} = 1$), a period of 44\,h, and the year 
2017. As the wave amplitude strongly depended on LT in 2017 
(section~\ref{sec:lines}), we only considered the four to five 1\,h intervals 
with the most reliable phases for each line, which essentially excluded the
evening data. In the case of 2019, six to eight intervals could be used. 
Intervals with less than nine 30\,min bins were never included. 
Line-dependent differences in the number of bins were only observed for 
intervals that were usually rejected. The optimum phase was calculated by 
weighted averaging using the inverse of the phase error of the fits of the 
selected LT intervals as weights (Table~\ref{tab:xs_phase}). Assuming the 
existence of a unique $\phi$ for the considered data, its uncertainty was 
derived from the weighted standard deviation of the phases of the individual 
fits. In the final step of the fit procedure, the hour-dependent fits were 
repeated with the phase fixed. In this way, the parameters $c$ and $a$ and 
their uncertainties were derived for each hour. 

The fitting of the SABER data could be simplified. The derivation of an
optimum phase was not necessary as only one LT range for 2017 provided 
reasonable fits due to the LT dependence of the wave amplitude (see 
section~\ref{sec:profiles}). For 2019, only one nighttime interval was
available (section~\ref{sec:saber}). As an alternative fitting approach for 
2017, we fitted all data simultaneously but with two additional fit 
parameters in order to scale $c$ and $a$ depending on the half of the night.
The results were almost identical but with slightly higher uncertainties. 
Therefore, we preferred the separate fits with the morning data as the 
reference for the phase derivation as in the case of the X-shooter data. In 
general, we fitted the vertically integrated VERs as well as the VER profiles 
from the two OH-related channels. For the latter, we first rebinned the 
profile data with a step size of 1\,km. Moreover, we usually started the 
fitting procedure at an altitude of 87\,km, where the OH emission is 
relatively strong. Thereafter, we used the resulting fit parameters as start 
values for the adjacent altitudes. The results for the latter were then taken 
for the next layers. In this way, we also obtained reasonable fits for 
heights with very weak OH emission (see section~\ref{sec:profiles}). 
Altitudes between 79 and 99\,km were considered.

\section{Results}\label{sec:results}

\subsection{OH Lines}\label{sec:lines}

\begin{figure}
\includegraphics[width=20pc]{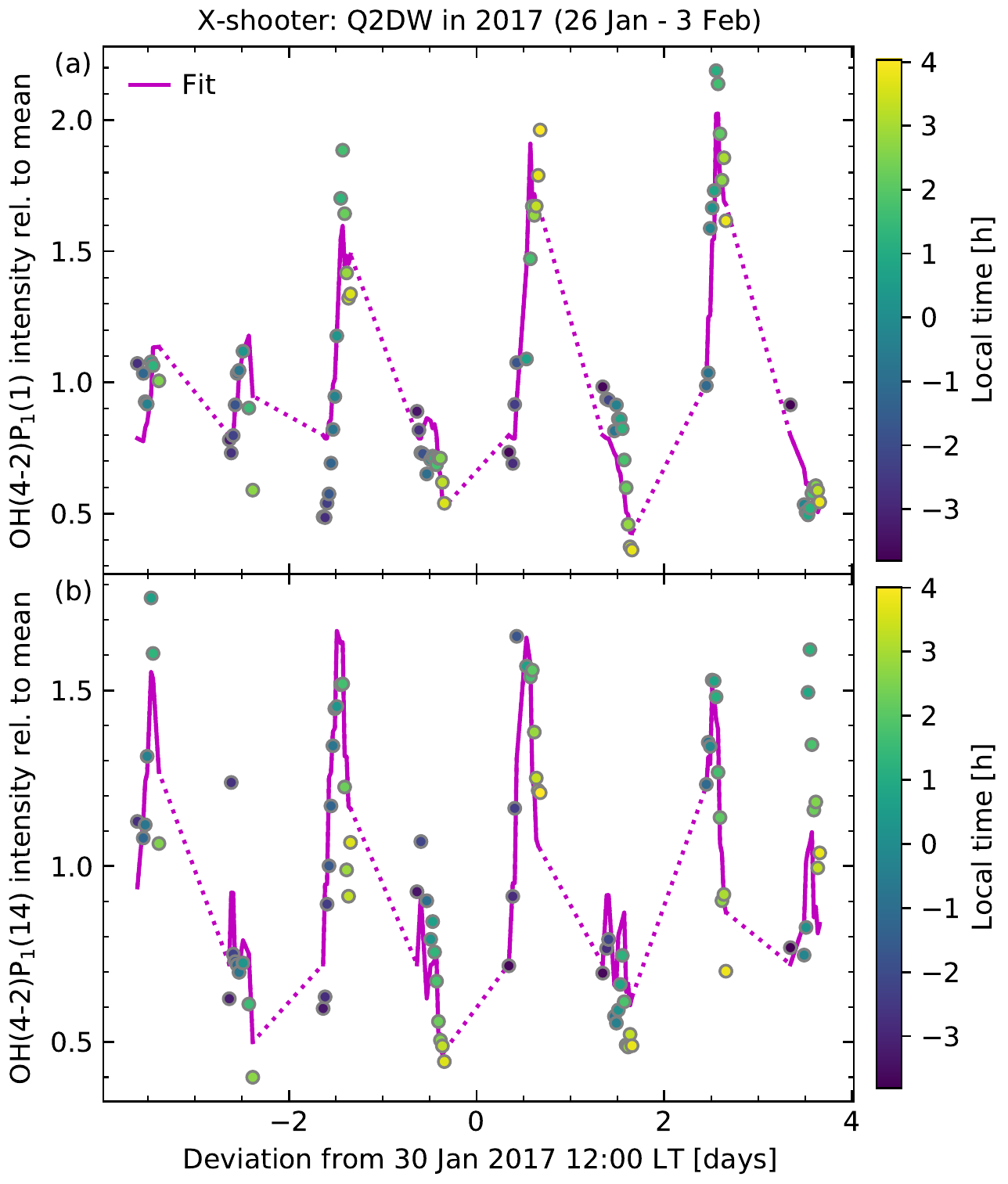}
\caption{X-shooter-based OH line intensity time series for the local time 
period from 26 January to 3 February 2017 given as deviation in days from 
mean solar noon at Cerro Paranal on 30 January. Circles show the intensities 
of OH(4-2)P$_1$(1) (a) and OH(4-2)P$_1$(14) (b) relative to the mean of the 
considered 30\,min bins. The local times of the latter are emphasized by 
different colors. The Q2DW fits with a fixed period of 44\,h, LT-dependent 
amplitudes, and line-specific phases are marked by solid lines. Daytime data 
gaps are bridged by dotted lines.}
\label{fig:xs_timeseries}
\end{figure}

In this section, we discuss the results of the Q2DW fits of the 
X-shooter-based OH line measurements for the wave event in 2017. 
Figure~\ref{fig:xs_timeseries} shows examples of the corresponding time 
series of 30\,min bins (section~\ref{sec:xshooter}) for the $\Lambda$ 
doublets OH(4-2)P$_1$(1) (a) and OH(4-2)P$_1$(14) (b), which only differ by 
the rotational quantum number. However, the difference in the rotational 
energy of 3,239\,cm$^{-1}$ is the largest of the entire set of 298 lines. 
Hence, the variability patterns should deviate. The inspection of the figure 
confirms this assumption. Deviations can be observed in the day-to-day 
development of the maximimum and minimum relative intensities and the local 
times showing these extreme values. Nevertheless, the basic pattern is 
similar. There are alternating nights with low and high relative intensities, 
which clearly points to the presence of a Q2DW. The apparent amplitudes are 
large. For OH(4-2)P$_1$(1), the maximum-to-minimum ratio is 6.1. The standard
deviation relative to the mean for the 88 bins indicates 0.45. For 
OH(4-2)P$_1$(14), the corresponding values for the 86 available bins are 
slightly lower (4.4 and 0.36, respectively).

\begin{figure}
\includegraphics[width=20pc]{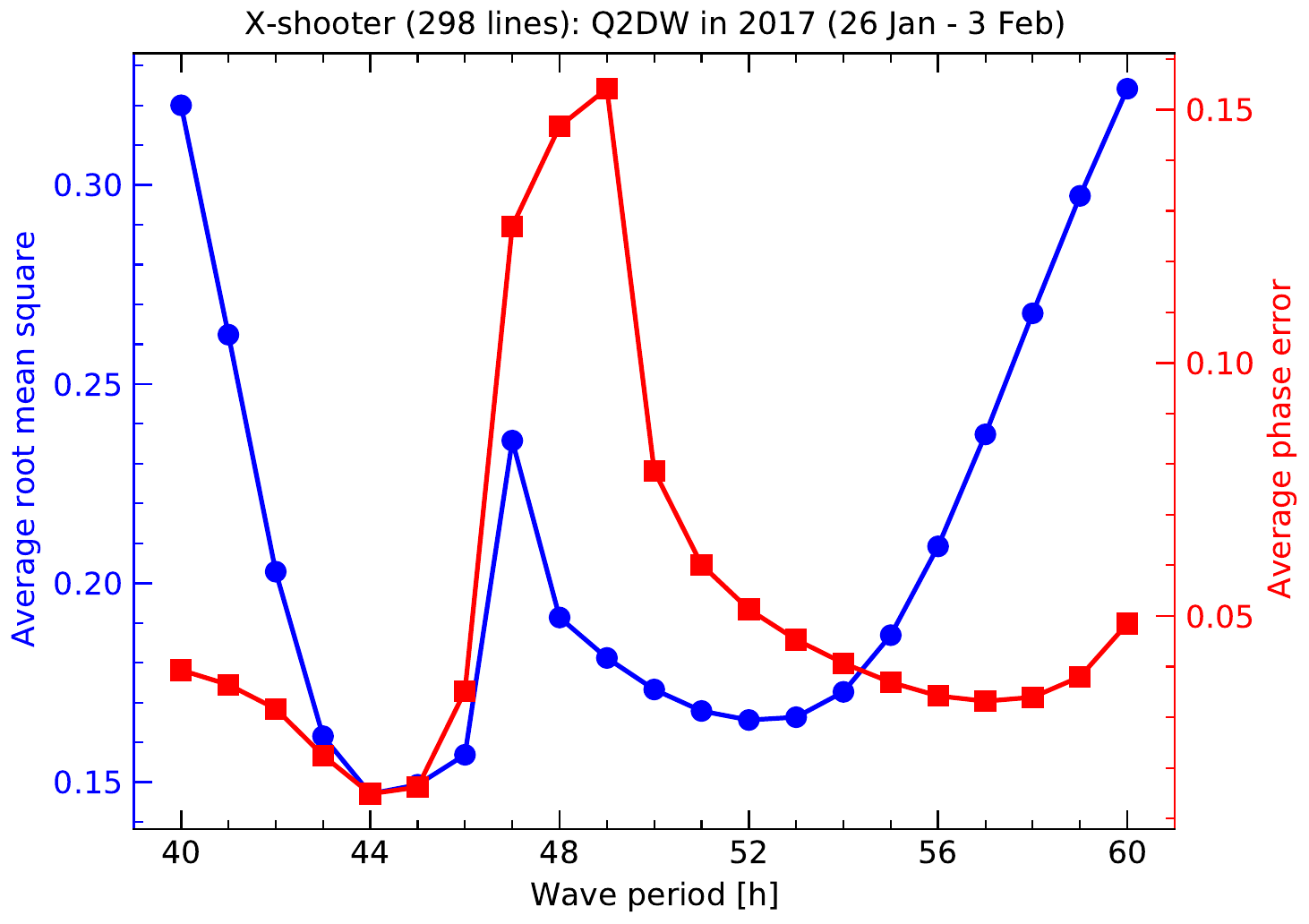}
\caption{Derivation of the most likely period for the Q2DW in 2017 based on 
the root mean square of the relative intensity fit (circles, left axis) and 
the phase error relative to the period (squares, right axis) averaged for all 
298 OH lines.}
\label{fig:xs_period}
\end{figure}

Figure~\ref{fig:xs_timeseries} also displays our best fits with LT-dependent
amplitudes for both OH emissions. The fits were performed for all LT bins 
except for 04:00 to 05:00, which includes only a single observation 
(Table~\ref{tab:xs_phase}). The fixed period for all LT bins was set to 
44\,h. This decision is justified by Figure~\ref{fig:xs_period}, which shows 
the root mean square (rms) for the differences between fit and observed data 
relative to the line-specific mean intensity averaged for all 298 lines as a 
function of the period grid from 40 to 60\,h. Moreover, the average phase 
error relative to the period (Table~\ref{tab:xs_phase}) is displayed for the 
same set of lines and periods. The minimum values of both indicators (0.15 
for the rms and 0.015 for the phase uncertainty) are clearly located at 
44\,h. The accuracy of this period, which is limited by the step size of 
1\,h, is highly sufficient for the estimation of emission heights as we will 
discuss in section~\ref{sec:heights}. Moreover, note that the scatter in the
periods from the minimum rms derived for the individual lines is fairly
small (63\% with 44\,h and 33\% with 45\,h) despite significant differences 
in the time series. Q2DW periods below 48\,h appear to be rather typical in 
regions close to Cerro Paranal as long-term OH observations at El Leoncito 
(31.8$^{\circ}$\,S, 69.3$^{\circ}$\,W) indicate \cite{reisin21}. 
Figure~\ref{fig:xs_period} reveals that periods close to 2 full days are not 
able to reproduce the change of the variability pattern from night to night 
shown in Figure~\ref{fig:xs_timeseries}. In particular, the small apparent 
amplitudes of the first two nights for OH(4-2)P$_1$(1) suggest that the 
maximum and minimum were outside the range of local times covered by the 
X-shooter data, i.e. too late in the morning. With a period distinctly 
smaller than 48\,h, it was then possible to observe the extreme intensities 
afterwards at nighttime. Therefore, the time series is just long enough to 
robustly derive a wave period. The variability pattern of OH(4-2)P$_1$(14) is 
more regular at the beginning with high deviations from the mean intensity, 
which excludes that the Q2DW was significantly weaker in the first nights as 
the consideration of only the P$_1$(1) data could suggest. The differences 
between both lines are obviously the result of shifts in the positions of the 
extremes. Consequently, both emissions need to show different Q2DW phases. 
Our fits revealed phases of 0.403 and 0.229 at the reference time for the 
lines with $N^{\prime} = 1$ and 14, respectively. The resulting phase 
difference is much larger than the mean uncertainty reported above (which 
should be even smaller for phase comparisons). Hence, the wave of 2017 is 
obviously suitable to safely separate lines based on their fitted phases.     
  
\begin{figure}
\includegraphics[width=20pc]{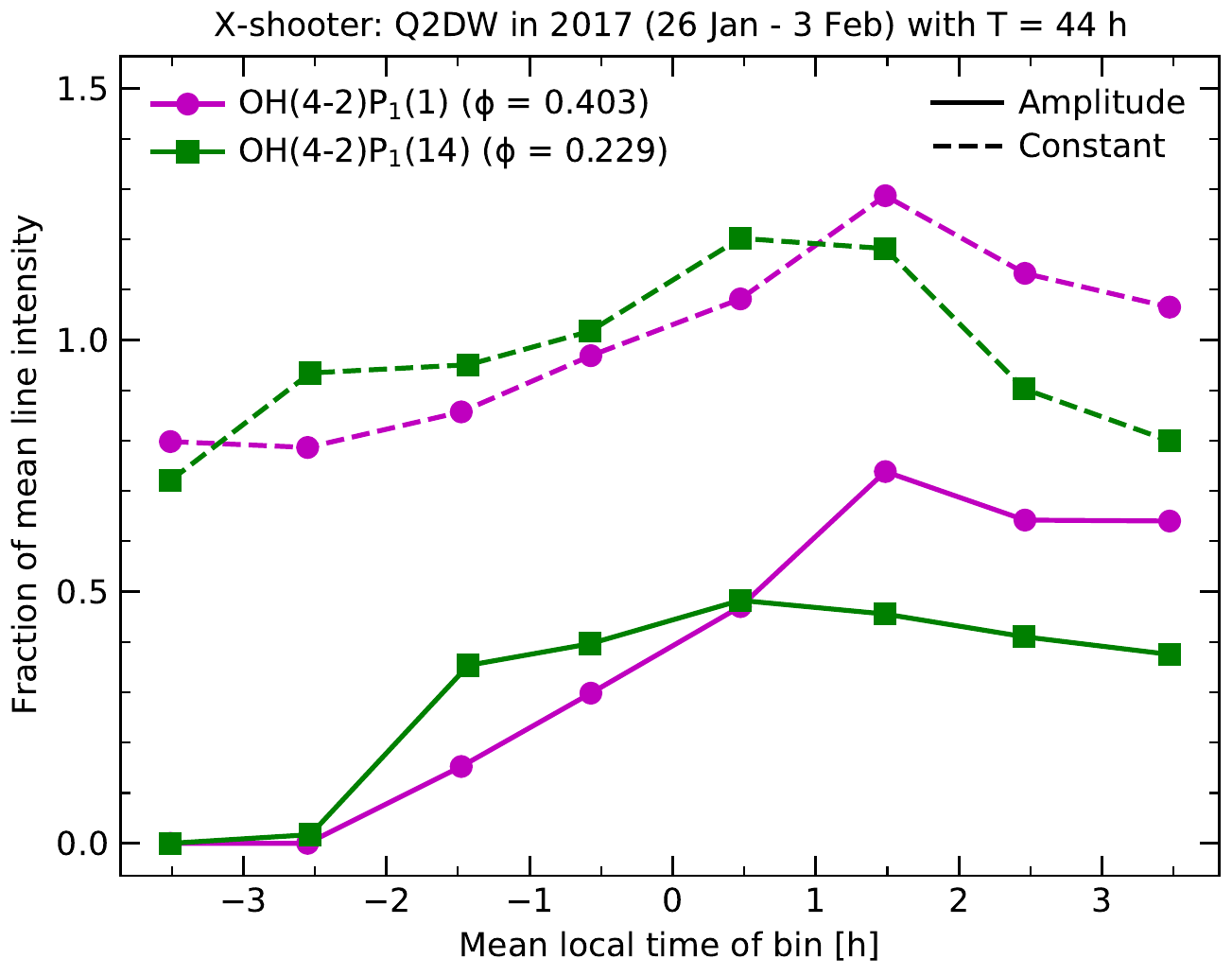}
\caption{Dependence of fitted Q2DW parameters on local time for 
OH(4-2)P$_1$(1) (circles) and OH(4-2)P$_1$(14) (squares). The abscissa shows
the mean local time for nighttime intervals with a width of 1\,h. Results are
provided for the amplitude $c{\cdot}a$ (solid lines) and the additive 
constant $c$ (dashed lines). In all cases, the plotted values are relative to 
the mean line intensity for the Q2DW-related sample. The mean fit 
uncertainties for $c{\cdot}a$ and $c$ for data points with non-zero weight 
for the phase derivation (Table~\ref{tab:xs_phase}) are 0.09 and 0.06, 
respectively. The legend also provides the effective phases $\phi$ relative 
to the period of 44\,h for both OH lines (cf. Table~\ref{tab:xs_phase}).}
\label{fig:xs_ampl_vs_hour}
\end{figure}

The results for the two example $\Lambda$ doublets are further analyzed in
Figure~\ref{fig:xs_ampl_vs_hour}, which shows the amplitude, i.e. the product
of the fit parameters $c$ and $a$ relative to the mean intensity of the time 
series, as a function of the mean local time for intervals with a width of 
1\,h. The wave amplitudes of both emissions strongly depend on local time. 
There is no detection of a Q2DW before 22:00. Afterwards the amplitudes 
increase and then become relatively stable. This development is faster for 
OH(4-2)P$_1$(14), where a shallow maximum with an amplitude of 0.46 is 
visible between 0:00 and 01:00. The maximum for OH(4-2)P$_1$(1) is 
significantly higher and amounts to 0.74. It is present between 01:00 and 
02:00. These remarkable structures cannot be explained by a single wave with 
fixed amplitude. A model with multiple wave components would certainly better 
reproduce the patterns, but at least a model with two components did not 
return reliable wave properties (section~\ref{sec:methods}). The persistent 
lack of wave-like variability at the beginning of the nights is hard to 
explain in this way. Hence, our assumption of LT-dependent Q2DW amplitudes,
which leads to the narrow peaks of the fit models in 
Figure~\ref{fig:xs_timeseries}, appears to be the most promising approach for 
achieving a satisfying agreement with the observed time series. This model 
requires that the physical conditions that control the wave propagation 
and/or the sensitivity of the OH emission for Q2DWs change depending on the 
time of the day. Nonlinear interaction with especially the diurnal and 
semidiurnal tides \cite<e.g.,>{palo99} could explain this behavior. In this 
context, phase locking, which would significantly weaken the diurnal tide 
\cite{walterscheid96,hecht10,walterscheid15}, did not seem to be active at
Cerro Paranal during the considered time interval as periods very close to 
48\,h cannot explain the time series. The LT dependence of the Q2DW amplitude 
in OH emission might also be affected by the negative nocturnal trend of 
atomic oxygen (produced at daytime) that would always be present without 
vertical dynamics especially at the lowest OH emission altitudes 
\cite{marsh06}. 

Figure~\ref{fig:xs_ampl_vs_hour} also shows the constant (or factor) $c$
relative to the mean intensity. For an undisturbed cosine oscillation 
centered on the mean intensity of the time series, $c$ should be close to 1 
(see section~\ref{sec:methods}). In reality, the values range between 0.7 and 
1.3. They cause that the maximum $c{\cdot}a$ are not located at the end of 
the night as in the case of $a$ (not shown). Although the joint fitting of 
$c$ and $a$ might lead to some systematics due to possible degeneracies, 
asymmetries in very large intensity variations that cause larger deviations 
during the crest of the wave are more likely. Moreover, there are also 
variations in the nocturnal OH intensity without a wave. An inspection of the 
entire X-shooter data set suggests that this is probably a minor effect as 
the climatological changes for the investigated times in southern summer are 
of the order of only 10\% \cite<cf.>{noll17}. The intensity even tends to 
decrease with increasing LT in contrast to our results for $c$.      

\begin{figure}
\includegraphics[width=20pc]{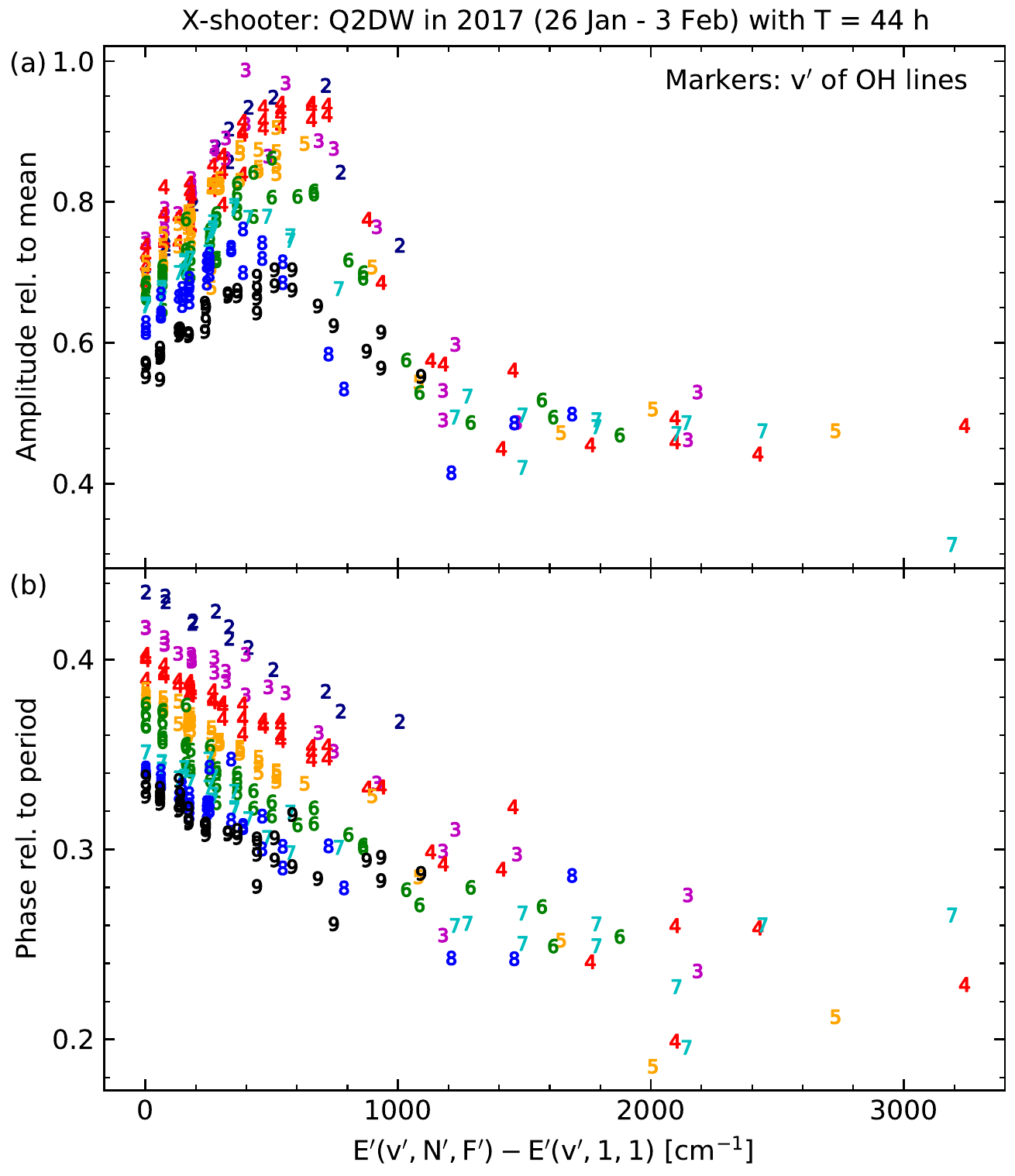}
\caption{Maximum amplitude $c{\cdot}a$ relative to the mean (a) and phase 
$\phi$ relative to the period at 30 January 12:00 LT at Cerro Paranal (b) for 
all 298 OH lines used for the fit of the Q2DW event in 2017. The abscissa 
shows the energy of the upper level of the transition minus the lowest energy 
for the corresponding vibrational state $v^{\prime}$. The latter is given by 
colored numbers. The representative amplitude (phase) uncertainties are about 
0.021 (0.005), 0.023 (0.008), and 0.038 (0.017) for energy differences below 
400\,cm$^{-1}$, between 400 and 800\,cm$^{-1}$, and above 800\,cm$^{-1}$, 
respectively.}
\label{fig:xs_ampl_phase}
\end{figure}

Next, we discuss the whole set of 298 lines. 
Figure~\ref{fig:xs_ampl_phase}a shows the line-specific maximum of the
amplitude $c{\cdot}a$ of all hour intervals as a function of the energy of 
the upper state of the transition relative to the lowest energy of the 
corresponding vibrational level $v^{\prime}$. The reference states are 
characterized by a rotational quantum number $N^{\prime} = 1$ and the 
electronic substate $\mathrm{X}^2\mathrm{\Pi}_{3/2}$ ($F^{\prime} = 1$)
\cite<e.g.,>{noll20}, i.e. the energy difference equals 0\,cm$^{-1}$ for 
P$_1(1)$ and Q$_1(1)$ lines. If we focus on these lines, the figure reveals a 
clear increase of the amplitude for decreasing $v^{\prime}$ with values 
between 0.56 for $v^{\prime} = 9$ and 0.74 for $v^{\prime} = 3$. This trend 
is highly significant as the uncertainties derived from lines with the same 
upper level are only about 0.02. The amplitude differences for adjacent 
vibrational levels are larger for high $v^{\prime}$. For increasing 
rotational energy, we find an increase for all vibrational levels which is 
more pronounced for low $v^{\prime}$. Up to about 500\,cm$^{-1}$, the 
increase ranges from 0.12 for $v^{\prime} = 9$ to 0.23 for $v^{\prime} = 2$. 
For the lowest vibrational levels, this rise results in enormous amplitudes 
of more than 90\% of the sample mean intensity. At higher rotational 
energies, the amplitudes decrease again. This trend appears to start slightly 
later for lower $v^{\prime}$, but not later than about 700\,cm$^{-1}$. The 
drop of the amplitude is much stronger for lower $v^{\prime}$. For the 
highest $N^{\prime}$, the amplitude appears to be independent of the 
vibrational level and is distinctly smaller than for low $N^{\prime}$. The 
mean amounts to about 0.48 for energy differences $\Delta E^{\prime}$ larger 
than 1300\,cm$^{-1}$. 

The complex pattern in Figure~\ref{fig:xs_ampl_phase}a needs an 
explanation. As we can expect that the emission altitudes increase with 
higher $v^{\prime}$ and $N^{\prime}$ 
\cite<e.g.,>{adler97,dodd94,noll18b,savigny12}, lower amplitudes with 
increasing quantum numbers imply a decrease of the wave amplitude with
increasing altitude. Such a trend is consistent with the dependence of the
amplitude on $v^{\prime}$ for $\Delta E^{\prime}$ up to at least 
1,000\,cm$^{-1}$ and the comparison of amplitudes for very low and very high 
$N^{\prime}$. An issue seems to be that the highest amplitudes are located at 
intermediate rotational energies. However, this feature can be explained by
means of the structure of the OH rotational level population, which can be
characterized by a cold and a hot population for each $v^{\prime}$ 
\cite{cosby07,kalogerakis18,kalogerakis19,noll20,oliva15}. While the 
temperature of the cold population, which dominates the emission at low
$N^{\prime}$, is consistent with the ambient temperature, i.e. about 190\,K
on average at Cerro Paranal \cite{noll16,noll20}, the hot population shows 
apparent temperatures between about 700\,K for $v^{\prime} = 9$ \cite{noll20}
and about 12,000\,K for $v^{\prime} = 2$ \cite{oliva15}. Such extreme values 
reflect significant deviations from the local thermodynamic equilibrium (LTE), 
which are related to the non-LTE nascent population of the hydrogen--ozone 
reaction \cite{llewellyn78} as well as an insufficient frequency of collisions 
that thermalize the rotational level population compared to $v$-changing 
collisions and the radiation of airglow photons \cite{noll18b}. An inspection 
of the combined populations now shows that the contributions of the cold and 
the hot component are of the same order at similar $\Delta E^{\prime}$ as 
those where we find the maximum amplitudes in 
Figure~\ref{fig:xs_ampl_phase}a \cite{noll20}. Consequently, the latter 
can be explained by an increased sensitivity of the corresponding OH lines to 
the Q2DW due to the strong impact of the wave-induced variation of the ambient 
temperature (affecting the cold population) on the rotational energy where 
cold and hot populations have similar contributions. Changes in the cold 
component are more crucial because of its much steeper decline with increasing 
energy. Also note that the relative contribution of the hot population is 
lower than a percent for $v^{\prime} \le 6$ at 0\,cm$^{-1}$ 
\cite{noll20,oliva15}. The rapid growth of this fraction with increasing 
$\Delta E^{\prime}$ can then explain the increase of the Q2DW amplitudes until
the maximum values at 500 to 700\,cm$^{-1}$. Moreover, the drop of the 
amplitudes marks the energies where the contribution of the cold population 
becomes minor. Hence, the highest $N^{\prime}$ show a pure hot population, 
which can obviously be described by a single amplitude. The lack of a 
dependence of the amplitude on $v^{\prime}$ suggests that the 
$v^{\prime}$-specific hot components are linked and might be represented by a 
single population (at least in part). The structure of the full 
roto-vibrational populations, which show connections between high $N^{\prime}$
for adjacent $v^{\prime}$ \cite{cosby07,noll20}, seems to confirm this 
interpretation.  

Q2DW phases relative to the period of 44\,h at 30 January 2017 12:00 LT are
shown for all 298 OH $\Lambda$ doublets in 
Figure~\ref{fig:xs_ampl_phase}b. In general, there is a clear decrease 
of the phase with increasing $v^{\prime}$ and $N^{\prime}$. For the levels
with $F^{\prime} = 1$ and $N^{\prime} = 1$, the phase ranges from 0.436 for
$v^{\prime} = 2$ to 0.333 for $v^{\prime} = 9$, which is distinctly more 
than the uncertainty of 0.005 derived from lines with the same upper levels. 
For higher $N^{\prime}$, the phases decrease almost linearly and similar for 
all $v^{\prime}$ with shifts of about 0.04 up to about 500\,cm$^{-1}$. The 
trend may even hold until about 1,500\,cm$^{-1}$ but with a decreasing 
difference between low and high $v^{\prime}$. For the highest $v^{\prime}$, 
there is a flattening of the phase change, which might even stop at the end. 
However, this remains questionable due to the relatively small number of 
lines and relatively high uncertainties. The average phase for 
$\Delta E^{\prime} \ge 2,000$\,cm$^{-1}$ is 0.23, which is about 0.2, i.e. 
almost 9\,h, lower than the maximum shown by OH(2-0)P$_1$(1). This relatively
large spread is promising in terms of a sensitive derivation of effective 
emission heights for various OH emission lines (see 
section~\ref{sec:heights}).

\subsection{OH Emission Profiles}\label{sec:profiles}

\begin{figure}
\includegraphics[width=20pc]{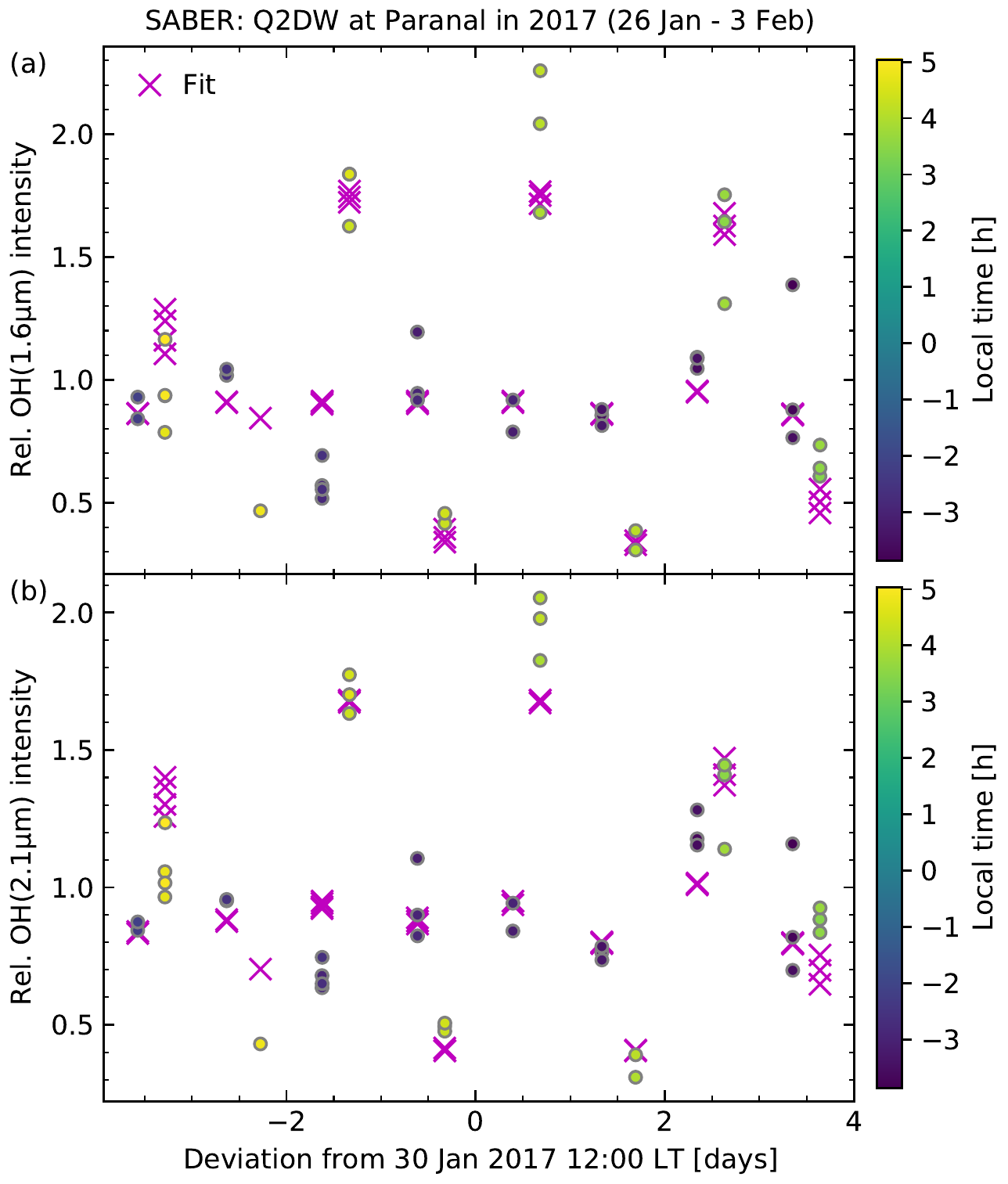}
\caption{SABER-based vertically integrated OH VERs for the local time period 
from 26 January to 3 February 2017 given as deviation in days from mean solar 
noon at Cerro Paranal on 30 January. The results for the OH-related channels 
centered on 1.6\,$\mu$m (a) and 2.1\,$\mu$m (b) are shown relative to the 
mean of the considered 44 data points (circles). The local times at the 
reference coordinates of the observations are emphasized by different colors. 
The Q2DW fits for the two LT ranges are marked by crosses.}
\label{fig:sb_timeseries}
\end{figure}

In order to link wave phases and heights, we need altitude-resolved OH 
emission measurements. The required data were obtained by the two OH channels 
of the limb-scanning SABER radiometer (section~\ref{sec:saber}). For a 
comparison with the X-shooter-based time series, 
Figure~\ref{fig:sb_timeseries} shows vertically integrated VERs of the 
channels OH(1.6\,$\mu$m) (a) and OH(2.1\,$\mu$m) (b) relative to the mean of 
the 44 measurements that are representative of the region around Cerro 
Paranal during the eight nights in 2017 covered by the X-shooter data set. 
Both channels show variability patterns that are consistent with a strong 
Q2DW for the 22 data points taken at about 04:00 LT, whereas the data related 
to about 21:00 LT do not display clear wave features. These results agree 
with those of the X-shooter data set in terms of the pronounced time 
dependence of the wave amplitude (Figure~\ref{fig:xs_ampl_vs_hour}). The 
maximum-to-minimun ratios and standard deviations relative to the mean are 
7.4 and 0.46 for all OH(1.6\,$\mu$m) data and 6.6 and 0.42 for all 
OH(2.1\,$\mu$m) data. The smaller values for the latter are consistent with 
the decrease of the wave amplitude with increasing $v^{\prime}$ for the 
X-shooter data (Figure~\ref{fig:xs_ampl_phase}). Remember that the effective 
$v^{\prime}$ are about 4.6 and 8.3 for the two SABER OH channels 
(section~\ref{sec:saber}). The maximum-to-minimum ratios are higher than the 
values of 6.1 for OH(4-2)P$_1$(1) and especially 4.4 for OH(4-2)P$_1$(14) 
(section~\ref{sec:lines}). The standard deviations are very similar to the 
value of 0.45 for the line with $N^{\prime} = 1$ but higher than the result
of 0.36 for the line with high $N^{\prime}$. These findings agree well as the 
integrated emission of the SABER channels is dominated by lines with small 
$N^{\prime}$. Consequently, SABER and X-shooter data show a consistent 
picture of the Q2DW event in 2017. Differences in the measurement approaches 
and sample properties do not appear to have a significant impact. 

Figure~\ref{fig:sb_timeseries} also shows our fits of the two LT ranges 
based on the approach described in section~\ref{sec:methods}. Compared to 
the 04:00 LT data that we exclusively used for the phase derivation, the
amplitude $c{\cdot}a$ for the 21:00 LT data was significantly smaller. The 
ratios for the 21:00 and 04:00 amplitudes were about 0.09 for OH(1.6\,$\mu$m)
and about 0.20 for OH(2.1\,$\mu$m). Thus, the fit of the evening measurements
shows only very small Q2DW-related variations. Moreover, the corresponding 
factors for $c$ were 0.86 and 0.87, i.e. the intensity level for the evening 
data was lower on average. For the subsequent analysis, we only focused on 
the morning data.

\begin{figure}
\includegraphics[width=20pc]{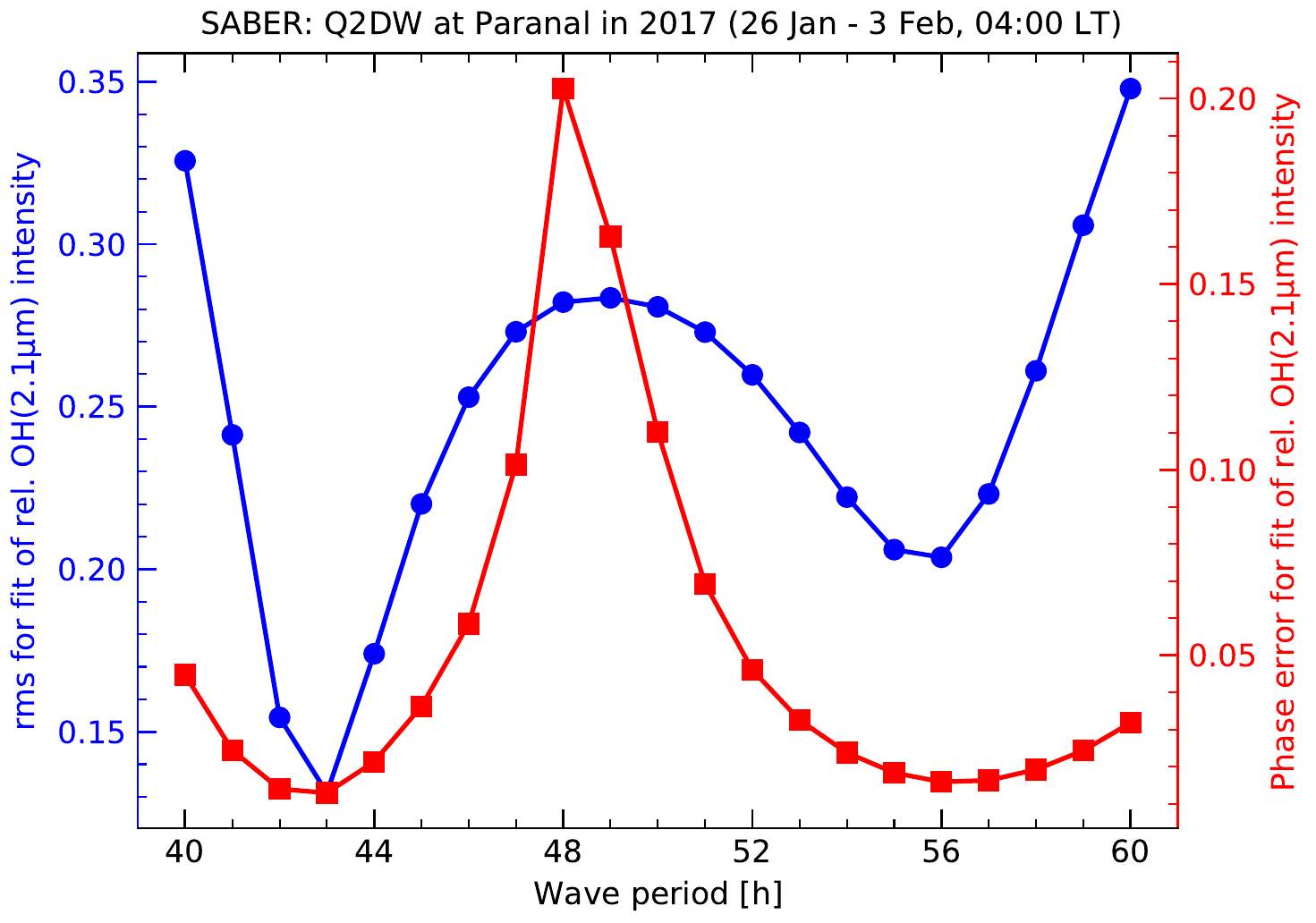}
\caption{Derivation of the most likely period for the morning data of the 
Q2DW in 2017 based on the root mean square of the relative intensity fit 
(circles, left axis) and the phase error relative to the period (squares, 
right axis) for the SABER OH channel centered on 2.1\,$\mu$m.}
\label{fig:sb_period}
\end{figure}

The fitting was performed for the same wave period as in the case of the
X-shooter data, i.e. 44\,h (section~\ref{sec:lines}). Using the same period
is important for the derivation of the effective emission heights. 
Nevertheless, we checked a wide range of periods in terms of the fit quality.
The results for OH(2.1\,$\mu$m) and the 04:00 LT data are shown in 
Figure~\ref{fig:sb_period}. The corresponding plot for OH(1.6\,$\mu$m) is 
very similar. The rms relative to the mean indicates a clear minimum of 0.13
at 43\,h. The phase error derived from the fit also shows a minimum there 
(0.013), although it is less pronounced. Consequently, the SABER-related fit 
returns the best result for a period which is only 1\,h lower than for the 
X-shooter data. This shift can also be observed for the maximum phase 
uncertainty (48 vs. 49\,h). In the view of the various differences between 
both data sets, the deviation is small. It is not crucial for the estimates 
of the line-specific emission altitudes. A comparison of the results for 43 
and 44\,h did not indicate significant deviations with respect to the general 
uncertainties (section~\ref{sec:heights}). In the following, we will focus on 
the results for 44\,h, which represents the best period from the X-shooter 
data set, which is much larger than the SABER data set in terms of OH 
emission features and observing times. 

Interestingly, the period can change significantly if the geographical area 
is not restricted to locations close to Cerro Paranal. Our additional 
check of the SABER data without longitude and LT limits (cf. 
section~\ref{sec:methods}) revealed a most likely period of 49\,h for the two 
OH channels. This result is in good agreement with the findings of 
\citeA{gu19}, who reported a period of 48\,h for the Q2DW in 2017 based on 
the kinetic temperatures from SABER. We repeated this fit and found the same. 
For the global fits, we preferentially used the combined evening and morning 
data, i.e. we neglected a possible LT dependence of the amplitude. This 
setting resulted in better fits, which is also in contrast to our experience 
with the data restricted to the region around Cerro Paranal. Moreover, an 
inspection of the wave properties depending on the longitude showed that the 
South American sector exhibited the clearest Q2DW variability pattern. In 
other regions, the wave was much less pronounced. This observation may 
explain why \citeA{gu19} only identified an intermediate wave amplitude of 
10\,K at 82.5\,km for the event in 2017. The latitude- and time-resolved
SABER-based results for a fixed period of 48\,h from \citeA{xiong18} show 
amplitudes slightly below 10\,K at 84\,km at about 25$^{\circ}$\,S for the 
relevant time interval at the end of January and beginning of February. We 
find about 18\,K around Cerro Paranal in the morning. In addition, the local
times with the strongest variations changed depending on longitude, which 
indicates why global fits without the corresponding scaling factors worked 
better. This complex variability pattern suggests that the Q2DW is 
significantly disturbed by interactions with other planetary waves, tides, 
gravity waves, and/or the mean flow. For OH emission from SABER observations,
\citeA{pedatella12} already found a complex longitude dependence of the 
Q2DW-related variability due to contributions of nonmigrating tides, 
stationary planetary waves, and secondary waves that are preferentially 
generated by the nonlinear interaction of the Q2DW and migrating tides. The 
latter supports our assumption that such interactions were important for the 
Q2DW event in 2017 with a pronounced LT dependence of the measured 
amplitude.            

\begin{figure}
\noindent\includegraphics[width=\textwidth]{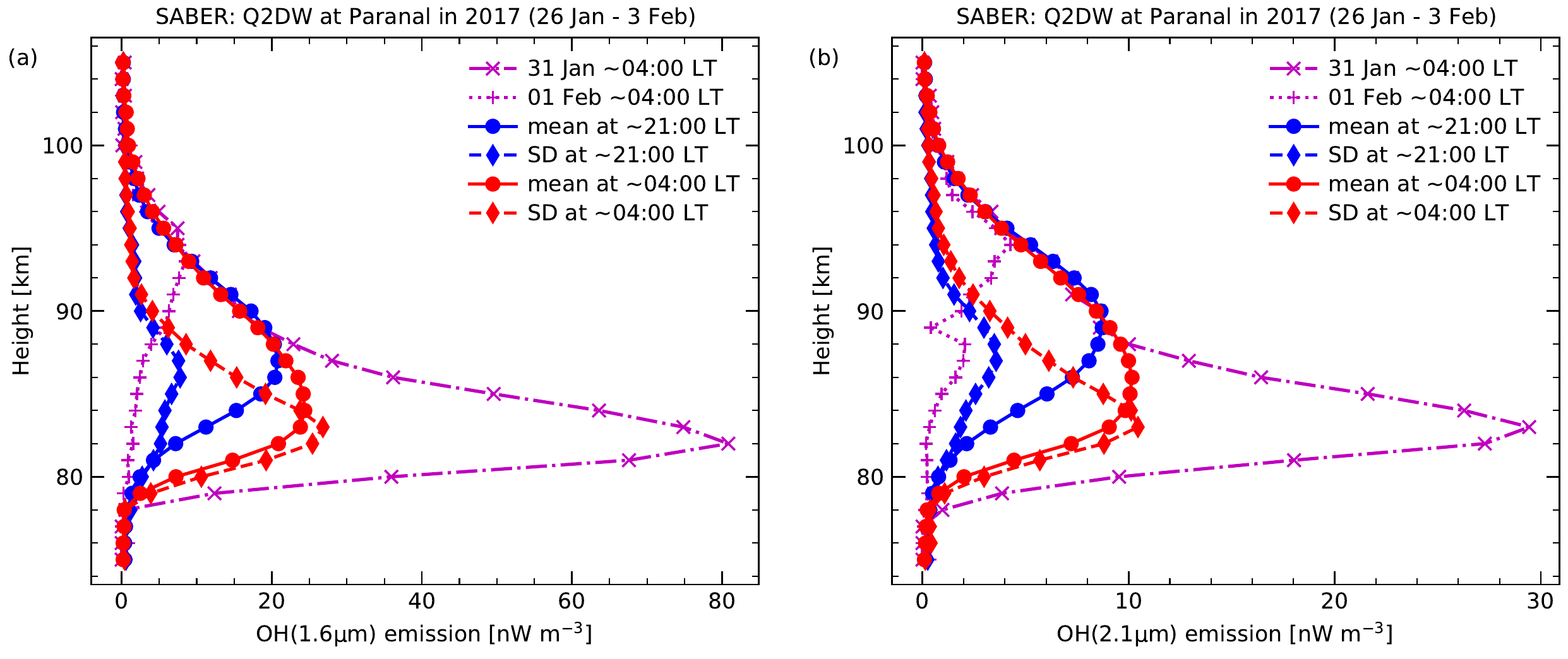}
\caption{Vertical VER profiles for the Q2DW event in 2017 from the SABER OH 
channels centered on 1.6\,$\mu$m (a) and 2.1\,$\mu$m (b). As described in the 
legend, both subfigures show two individual profiles with dates separated by 
about 24\,h (crosses and plus signs), mean profiles (circles) for the 
evening (upper peak) and morning (lower peak), as well as standard deviation 
(SD) profiles (diamonds) for the evening (upper peak) and morning (lower 
peak). For the calculation of the curves representing the evening data, one 
profile with unusually large VERs below 80\,km had been excluded.}
\label{fig:sb_profiles}
\end{figure}

We now turn to the discussion of the height-dependent impact of the Q2DW.
Figure~\ref{fig:sb_profiles} shows different kinds of emission profiles for
the two OH channels centered on 1.6\,$\mu$m (a) and 2.1\,$\mu$m (b). For each
channel, two individual profiles from 31 January and 1 February 2017 at 
about 21:00 LT are displayed. The huge differences between both profiles 
with a time difference of 1 day demonstrate the implications for the OH
emission by a large amplitude Q2DW. On 31 January, the OH(1.6\,$\mu$m) 
profile peaked at a very low altitude of 82\,km with an enormous VER of 
81\,nW\,m$^{-3}$. In contrast, the emission almost vanished completely
at this altitude on the next day. As only the emission at the highest 
altitudes remained relatively similar, the peak moved upward by 11\,km 
to 93\,km, where the VER was only 8\,nW\,m$^{-3}$. The situation for 
OH(2.1\,$\mu$m) is very similar with the peaks just 1\,km higher. It is not 
clear whether the VER minimum at 89\,km on 1 February is real as it cannot be 
seen in the corresponding data from OH(1.6\,$\mu$m).  

Figure~\ref{fig:sb_profiles} also shows mean and standard deviation profiles
for the two LT regimes of about 21:00 and 04:00 in the eight selected nights.
The evening data indicate typical mean profiles with respect to the long-term
averages for Cerro Paranal \cite{noll17}. The centroid altitudes of the 
plotted profiles of both channels are 88.1 and 89.5\,km, respectively. The 
standard deviation profiles of the 21:00 LT data only indicate relatively 
weak variations. The peaks of these profiles are about 1 to 2\,km lower than 
for the mean profiles. These shifts are expected due to the steepening of the
atomic oxygen gradient with decreasing height \cite<e.g.,>{smith10}, which 
makes the lower altitudes more sensitive to vertical transport of this 
important gas for the OH production and subsequent relaxation and destruction
processes. The strong oxygen-induced emission variations at low altitudes 
lead to a significant perturbation of the mean profile for the morning data. 
The latter therefore extends to lower heights (with a 3\,km lower peak) and 
gets more broadened for both channels. The peak of the standard deviation 
profile moves in a similar way. The emission variability is much larger than 
in the case of the evening data (as expected). Note that the mean centroid 
emission altitudes of the individual profiles for 04:00 LT of 88.1\,km for 
OH(1.6\,$\mu$m) and 89.1\,km for OH(2.1\,$\mu$m) are very similar to those of
the 21:00 LT data (88.0\,km and 89.5\,km) since the large differences in the
VERs which affect the plotted mean profiles do not matter here. Nevertheless,
the larger impact of the Q2DW on the morning data can be recognized by the
standard deviation of the individual centroid altitudes of about 2.3\,km for
both OH channels, which is distinctly higher than about 0.6\,km for the 
evening data (if one outlier is excluded).
      
\begin{figure}
\includegraphics[width=20pc]{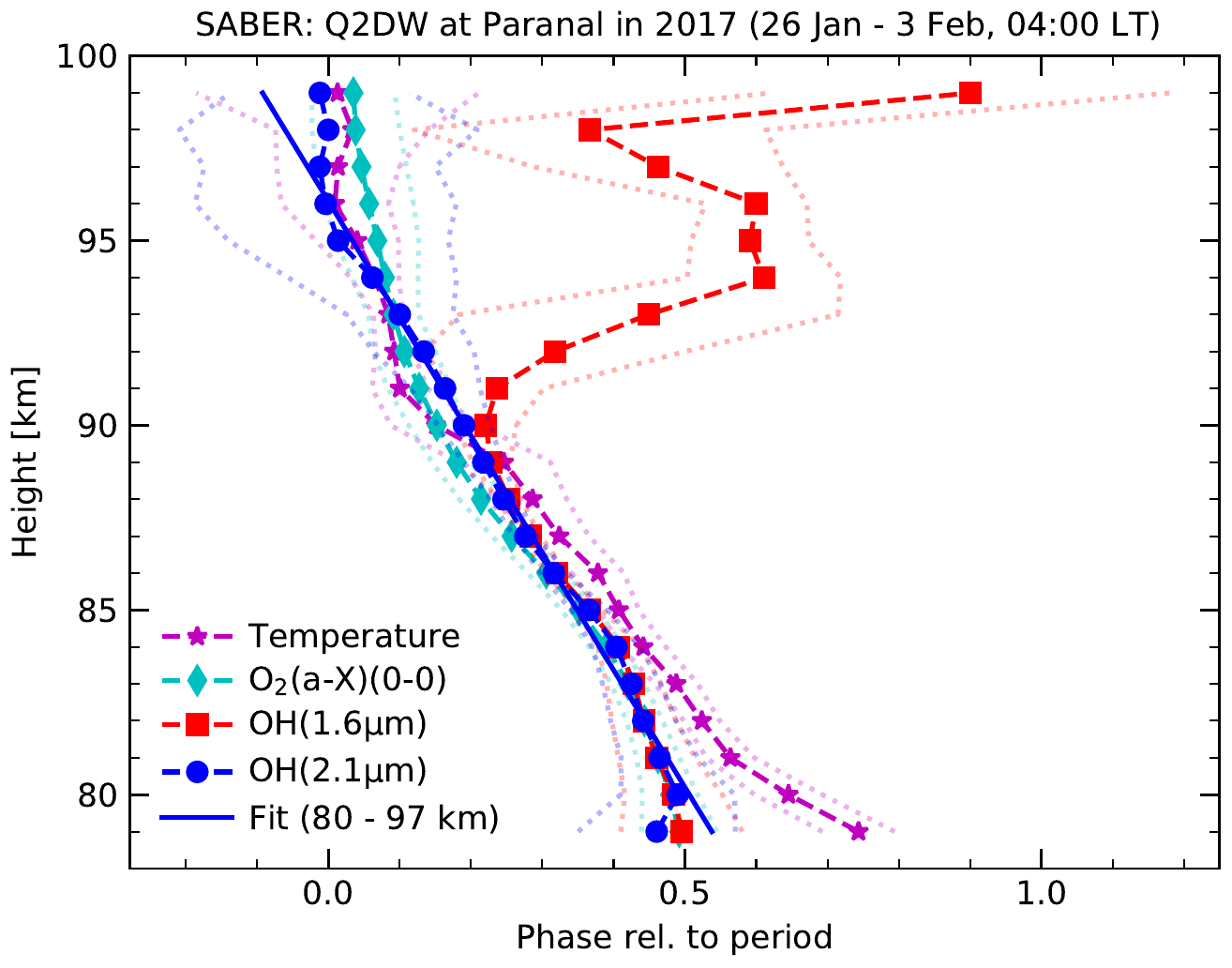}
\caption{Fitted phase relative to the period of 44\,h as a function of 
height for the SABER product kinetic temperature (stars) and the VERs of 
O$_2$(a-X)(0-0) at 1.27\,$\mu$m (diamonds), OH(1.6\,$\mu$m) (squares), and 
OH(2.1\,$\mu$m) (circles). The time series comprised the 22 limb scans taken
at about 04:00 LT in eight nights of 2017. The reference time for the plotted 
phases was 30 January 12:00 LT at Cerro Paranal. Fit uncertainties, which are
useful for relative quality comparisons, are indicated by dotted lines. 
Moreover, the plot shows the resulting regression line for a linear fit of 
the OH(2.1\,$\mu$m) phases in the altitude range between 80 and 97\,km.}
\label{fig:sb_phases}
\end{figure}

With the knowledge of the height-dependent response of the OH emission on the
passing Q2DW, we now discuss the wave phases as a function of altitude.  
Figure~\ref{fig:sb_phases} shows the phase functions between 79 and 99\,km 
from fits of the morning data with a period of 44\,h as described in 
section~\ref{sec:methods} for both OH channels. The phases are given for 30
January 2017 12:00 LT at Cerro Paranal. At least for the altitude range 
between 80 and 90\,km, there is a clear trend of decreasing phase with 
increasing height for the OH emissions at about 1.6 and 2.1\,$\mu$m. For 80 
to 89\,km, both curves are almost identical with a mean absolute difference 
of only 0.005. However, the discrepancy is rapidly growing above 89\,km. 
While the trend continues for OH(2.1\,$\mu$m), OH(1.6\,$\mu$m) shows a 
complex behavior with increasing and decreasing phase. The latter is not 
trustworthy. As expected from the ratio of standard deviation to mean 
profiles in Figure~\ref{fig:sb_profiles}, the fitted wave amplitudes relative 
to the mean rapidly decrease with increasing altitude. From 80 to 93\,km, 
$c{\cdot}a$ drops from 1.18 to 0.19 for OH(2.1\,$\mu$m). However, this is 
still moderate compared to a decrease from 1.23 to 0.04 for OH(1.6\,$\mu$m). 
Consequently, it appears that the wave-induced variability in the emission at 
about 1.6\,$\mu$m was too low to track the wave in a reliable way (cf. 
evening data in Table~\ref{tab:xs_phase} and 
Figure~\ref{fig:xs_ampl_vs_hour}), whereas it was still sufficient for 
OH(2.1\,$\mu$m). Note that the latter emission peaks higher in the atmosphere 
(Figure~\ref{fig:sb_profiles}). 

In order to check this interpretation, we also fitted the emission of 
O$_2$(a-X)(0-0) at about 1.27\,$\mu$m, which is also observed by SABER. The 
profile of this airglow emission strongly changes during the night 
\cite{noll16} due to the decay of an excited population originating from 
ozone photolysis at daytime \cite<e.g.,>{lopez89}. However, in the second 
half of the night, the emission distribution is similar to the one of OH. The 
fits only show a relatively weak decrease of the wave amplitude by a factor 2 
over the entire plotted altitude range. Hence, the resulting phases in 
Figure~\ref{fig:sb_phases} appear to be reliable. They clearly support the 
trend found for OH(2.1\,$\mu$m). This result is further confirmed by fits of
SABER-based kinetic temperature profiles \cite{dawkins18}, where the 
corresponding phase profile is also plotted. The temperature fits are 
relatively robust since the wave amplitude remained relatively high in the 
entire studied altitude regime (at least 11\,K up to 98\,km with maximum 
values of about 18\,K at 82\,km and 21\,K at 94\,km). The shape of the atomic 
oxygen profile does not matter for the temperature.
  
In conclusion, we have strong hints for a monotonically decreasing phase with
increasing height over the entire altitude range relevant for OH. This 
implication is ideal for the estimate of effective emission heights. 
Moreover, the earlier maxima at higher altitudes indicate that the wave was 
rising. This interpretation is also supported by test fits of the diurnal 
tide with the same fitting algorithm \cite<cf., e.g.,>{griffith21}. 
Consequently, the wave propagation is consistent with the assumed origin of
Q2DWs in the lower mesosphere \cite<e.g.,>{ern13}. The results suggest that 
the OH-relevant wave phases are best described by the profile fits for 
OH(2.1\,$\mu$m). There is an almost linear relation between phase and height 
over a wide altitude range. We found that the optimimum interval for a linear
regression extends from 80 to 97\,km. Then, we obtain a very high coefficent 
of determination r$^2$ of 0.995. The inverse slope of the regression line 
corresponds to the vertical wavelength $\lambda_\mathrm{z}$ of the wave. It 
amounts to $31.7 \pm 0.6$\,km, which is near the peak of the distribution of 
W3 Q2DW wavelengths derived by \citeA{huang13} based on SABER data. If we 
only consider the altitude range up to 89\,km, where OH(1.6\,$\mu$m) can also
be used, it is almost the same $\lambda_\mathrm{z}$ but with a larger 
uncertainty ($31.9 \pm 1.6$\,km). For OH(1.6\,$\mu$m), we then obtain 
$33.6 \pm 1.8$\,km, which agrees within the regression uncertainties.

\subsection{Effective OH Emission Heights}\label{sec:heights}

Combining the line-specific effective phases from section~\ref{sec:lines} 
with the slope of the relation between phase and height for OH(2.1\,$\mu$m)
from section~\ref{sec:profiles} for the same wave period of 44\,h allowed us 
to estimate the effective emission heights of 298 OH lines. For a direct
conversion, it is just necessary to subtract the phase for each line from
the intercept at 0\,km of 3.027 and to multiply the result with the vertical 
wavelength of 31.74\,km (see section~\ref{sec:profiles}). However, our 
calculation was more complex as we also considered possible phase deviations 
due to the differences in the lines of sight as well as geographical and 
temporal distributions of the X-shooter and SABER measurements for the Q2DW 
event in 2017. In order to estimate this effect, we used the fitted effective
phases of 0.388 and 0.345 for the vertically integrated VERs of the OH 
channels centered on 1.6 and 2.1\,$\mu$m, respectively, and compared them 
with X-shooter-based effective phases for sets of OH lines that are 
representative of these channels. We weighted the phases for individual lines
as shown in Figure~\ref{fig:xs_ampl_phase}b by the product of the measured 
line intensity and the channel-specific transmission \cite{baker07} at the 
line position. Moreover, we checked the impact of the fact that our line 
sample is not complete (section~\ref{sec:xshooter}). We found that the 
effective $v^{\prime}$ of our line mixes did not change for OH(1.6\,$\mu$m) 
but deviated by $+0.08$ for OH(2.1\,$\mu$m) compared to the reference values 
of 4.57 and 8.29 by \citeA{noll16}. However, lowering the contribution of the 
$v^{\prime} = 9$ lines to get a match in the effective $v^{\prime}$ for 
OH(2.1\,$\mu$m), i.e. 8.29, did not affect the effective phases, which turned
out to be 0.379 and 0.327 for the X-shooter data. Consequently, the SABER 
phases appear to be shifted by about 0.0135 on average. We subtracted this 
value from the intercept of the regression line before we calculated the 
effective emission heights. In terms of altitude, this shift corresponds to a
change of $-0.43$\,km with an uncertainty of 0.13\,km derived from the 
difference between the results for both channels.

For the estimate of the height uncertainties, we also checked whether our 
simple linear regression analysis without the consideration of the
height-dependent fit quality (Figure~\ref{fig:sb_phases}) could cause 
systematic phase offsets. For this purpose, we calculated a residual phase 
deviation from the differences between measured phase and regression line
at all heights weighted by the wave-induced variability. The resulting 
height uncertainty only amounts to 0.11\,km thanks to the convincing fit of
the OH(2.1\,$\mu$m) profile data. Concerning the X-shooter-related 
uncertainties, we used the representative phase uncertainties 0.005, 0.008,
and 0.017 (cf. caption of Figure~\ref{fig:xs_ampl_phase}), which are based on 
phase differences for OH lines with the same upper level for the energy
differences $\Delta E^{\prime}$ below 400\,cm$^{-1}$, between 400 and 
800\,cm$^{-1}$, and above 800\,cm$^{-1}$, respectively. Combined with the 
small uncertainties reported above, the effective absolute height errors 
resulted in about 0.24, 0.30, and 0.55\,km, respectively.
 
The effective heights for the 298 OH $\Lambda$ doublets range from 81.8 to
89.7\,km with an average of 84.88\,km and a standard deviation of 1.43\,km.
Compared to the uncertainties, the line-specific differences are highly
significant. The given altitudes are representative of the maximum emission 
variability induced by the Q2DW in the eight investigated nights of 2017. 
This can be illustrated for the vertically integrated VERs of the SABER 
channels at 1.6 and 2.1\,$\mu$m, for which we derived effective 
heights of 83.75 and 85.13\,km based on the already stated phases. The
profile plots for about 04:00 LT in Figure~\ref{fig:sb_profiles} indicate
that these wave-related heights are between the peak at 83\,km and the 
centroid altitude at 84.6 and 85.6\,km of the standard deviation profiles of 
the two channels. 

If the evening data of both instruments had resulted in reliable wave fits, 
we would probably have obtained significantly higher altitudes in agreement 
with the standard deviation profiles of these data. Hence, the resulting 
heights of our approach depend on the wave properties and the state of the 
background atmosphere during the analyzed time interval. Therefore, it is 
desirable to also provide heights for the studied OH lines which are 
representative of longer time scales. Moreover, there should be a closer
relation to the peak or centroid emission altitudes that are usually used in 
the literature, especially with respect to studies of OH rotational 
temperatures as indicators of the effective ambient temperature for the OH 
emission layer. The relevant heights for temperature and intensity variations
can differ by several kilometers \cite<e.g.,>{swenson98}. For example, we 
find for the ratio of the intensities of OH(3-1)P$_1$(1) and OH(3-1)P$_2$(2), 
which are often used for rotational temperature studies 
\cite<e.g.,>{beig03,schmidt13,noll15}, a phase shift of $-0.13$ compared to 
the mean phase for both lines that corresponds to an altitude difference of 
$+3.8$\,km with an uncertainty of several hundred meters (including a 
possible small discrepancy in the phase--height relations of temperature and 
OH intensity as indicated by Figure~\ref{fig:sb_phases}). Note that the
general use of temperatures instead of intensities for height estimates is
not possible as rotational temperatures show much higher measurement 
uncertainties and could vary in a different way than the satellite-based 
kinetic temperatures due to the non-LTE contributions which increase with 
increasing $v^{\prime}$ and $N^{\prime}$ \cite{noll20}. For the derivation 
of the reference heights, we considered the 14-year averages of the centroid 
altitudes of $87.81 \pm 0.02$\,km for OH(1.6\,$\mu$m) and 
$89.20 \pm 0.02$\,km for OH(2.1\,$\mu$m) from \citeA{noll17} that were 
calculated based on 4,496 SABER profiles taken close to Cerro Paranal. These 
values are about 4.06 and 4.07\,km higher than our results for the Q2DW phase
fits but they are close to the averages of the individual centroid altitudes
during the investigated event (section~\ref{sec:profiles}). The large 
difference can therefore be mainly explained by the strongly increasing VERs
for emission profiles with lower peaks and the resulting impact on the
vertical variability distribution. It is promising that the shifts are almost
the same for both OH channels as it implies that the Q2DW-based effective 
height differences between lines are also representative of the long-term 
averages of the centroid emission heights. At least for lines with low 
$N^{\prime}$ that mainly contribute to the SABER VERs, the deviations should 
be much smaller than the stated uncertainties of a few hundred meters. 
Consequently, we can also provide effective OH emission heights for average 
conditions at Cerro Paranal by shifting all line-specific heights by 
$+4.07$\,km. The resulting mean height for all 298 $\Lambda$ doublets is 
88.95\,km, which is in between the centroid altitudes for the reference 
profiles of both SABER OH channels. 

Our height estimates are based on a model that assumes a single wave with
a period of 44\,h and an amplitude depending on local time. In order to
increase the confidence in the corresponding results, it is important to 
know how changes in this model affect the derived emission heights. As 
already mentioned in section~\ref{sec:profiles}, we could have also used a
period of 43\,h as indicated by the SABER-based fit results in 
Figure~\ref{fig:sb_period}. In comparison to 44\,h, we obtained a mean 
height for the 298 lines which is 0.09\,km higher before and 0.02\,km lower 
after the shift. The standard deviation did not change, i.e. it is 1.43\,km.
Hence, the impact of a period change by 1\,h is negligible. We also 
performed a test for an extremely different period of 56\,h, which marks a 
secondary minimum of the phase error in Figure~\ref{fig:sb_period} and 
results in a downward-propagating Q2DW with a $\lambda_z$ of 
$53.9 \pm 2.3$\,km for OH(2.1\,$\mu$m). Nevertheless, the changes of the two 
mean heights were only $+0.21$ and $-0.17$\,km and the standard deviation 
just decreased by $0.08$\,km. This promising result demonstrates that even an 
unrealistic period can lead to reliable heights as long as a sufficiently 
linear phase--height relation with a significant spread of phases is present. 
For this reason, periods around 50\,h do not work for the analyzed 
Q2DW as they mark the reversal of the vertical propagation direction, which 
leads to very long $\lambda_z$. In section~\ref{sec:methods}, we have already 
discussed a two-wave model with fixed amplitudes that we checked as an 
alternative but resulted in unlikely wave parameters and low-quality fits. 
Nevertheless, even such a model appears to provide useful information on the 
OH height distribution. Focusing on the standard deviation of the effective 
heights of all lines, the best-fitting waves with periods of 43 and 51\,h and
opposite vertical propagation directions returned 2.18 and 0.72\,km. Both 
values show a large discrepancy but the average is 1.45\,km, which is very 
close to the value for our preferred model of 1.43\,km.
         
\begin{figure}
\includegraphics[width=20pc]{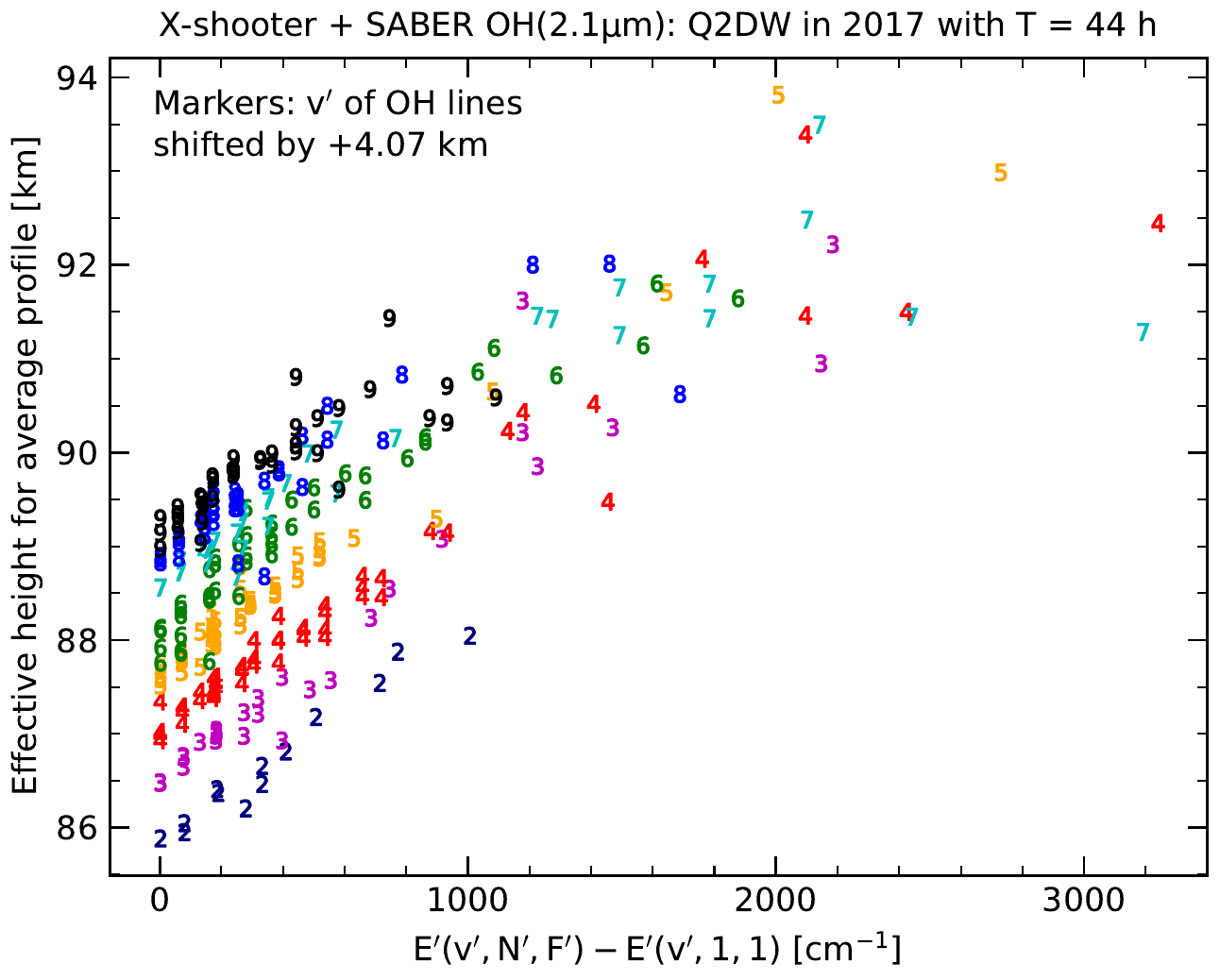}
\caption{Final effective heights for the considered 298 OH lines derived 
from a combination of fits of X-shooter line measurements and SABER VER data
of the OH channel centered on 2.1\,$\mu$m for the Q2DW event in 2017. The
wave-related effective heights were shifted upward by 4.07\,km to be 
representative of the mean SABER-related centroid altitudes for Cerro Paranal 
from \citeA{noll17}. The abscissa shows the energy of the upper level of the 
transition minus the lowest energy for the corresponding vibrational state 
$v^{\prime}$. The latter is given by colored numbers. The representative 
height uncertainties are about 0.24, 0.30, and 0.55\,km for energy 
differences below 400\,cm$^{-1}$, between 400 and 800\,cm$^{-1}$, and above 
800\,cm$^{-1}$, respectively.}
\label{fig:xssb_heights}
\end{figure}
 
With the confirmation of the robustness of our results, we now show the 
distribution of the reference heights for all investigated OH lines in
Figure~\ref{fig:xssb_heights}. The altitudes range from 85.9 to 93.8\,km, 
i.e. the maximum difference is almost 8\,km and therefore of the same order 
as the width of the full OH emission layer \cite<e.g.,>{baker88}. The 
detailed structure of the plotted data distribution was already discussed in 
terms of the wave phases in section~\ref{sec:xshooter}. The distribution is 
just inverted compared to the phases in Figure~\ref{fig:xs_ampl_phase}b. 
The effective emission heights increase for higher vibrational and rotational 
excitations as expected. Focusing on the levels with $F^{\prime} = 1$ and 
$N^{\prime} = 1$, the altitude difference between $v^{\prime} = 9$ 
(89.14\,km) and 2 (85.89\,km) amounts to 3.26\,km, which is about 0.47\,km 
for a difference of 1 in $v^{\prime}$ on average. This result agrees very 
well with the corresponding values found in previous studies of a few OH 
bands related to satellite data \cite{noll16,sheese14,savigny13,savigny12} 
and ground-based data combined with a sodium lidar \cite{schmidt18}. The 
height difference for $\Delta v^{\prime} = 1$ decreases with increasing 
vibrational excitation. Our analysis revealed 0.59\,km for 
$v^{\prime} \le 4$ but 0.29\,km (i.e. the half) for $v^{\prime} \ge 7$. This 
behavior agrees qualitatively with the modeling results of \citeA{savigny12} 
and is obviously caused by the $v^{\prime}$-dependent Einstein and 
collisional rate coefficients. As the previously published results refer to 
emissions of bands instead of lines, we also checked the change of the height
differences with increasing rotational energy. Taking only data with 
$\Delta E^{\prime}$ between 400 and 600\,cm$^{-1}$ ($N^{\prime}$ between 4 
and 6) as in section~\ref{sec:lines}, we found almost the same difference 
between $v^{\prime} = 9$ and 2 (3.20\,km) but at altitudes that are about 
1.1\,km higher. The change of the differences for $\Delta v^{\prime} = 1$ 
with $v^{\prime}$ also agrees within the uncertainties. Consequently, the 
results for entire bands should agree with those of the significantly 
contributing brighter lines of low to intermediate $N^{\prime}$. 

For the highest rotational levels, we do not see a clear dependence of the 
effective emission altitudes on $v^{\prime}$. The differences appear to
decrease with increasing $N^{\prime}$. As the nearly linear height increase
with $\Delta E^{\prime}$ seems to flatten (especially for higher 
$v^{\prime}$), there might even be a nearly constant effective emission 
height for the highest $N^{\prime}$. This behavior was already modeled by
\citeA{dodd94}. However, quantitatively, there are some discrepancies.      
Our mean for all lines with $\Delta E^{\prime} \ge 2,000$\,cm$^{-1}$ is 
92.3\,km. If we assume that this height is representative of all 
$v^{\prime}$, we estimate $v^{\prime}$-specific maximum changes compared to 
$\Delta E^{\prime} = 0$\,cm$^{-1}$ between 3.2 and 6.4\,km. \citeA{dodd94} 
only modeled 0 to 2\,km and a reference height for the highest $N^{\prime}$ 
of 89\,km. Based on the data set of \citeA{noll17} that we also used for the
derivation of our reference heights, \citeA{noll18b} modeled rotational level 
populations with a focus on $v^{\prime} = 9$. For two models with very 
different rate coefficients for collisions of OH with atomic oxygen, the 
maximum height changes resulted in 1.5 and 2.8\,km and a maximum emission 
height of almost 92\,km in the latter case. If we estimate the height changes
from the plotted data for $v^{\prime} = 9$, we obtain a rough lower limit of 
1.5 to 2\,km. These values might be better for a comparison since 
$v^{\prime} = 9$ levels with $\Delta E^{\prime} \ge 2,000$\,cm$^{-1}$ would 
be far beyond the exothermicity limit of the hydrogen--ozone reaction 
\cite{cosby07,noll18b} of about 1,070\,cm$^{-1}$. In any case, the model 
results appear to match the correct order of magnitude, although the 
uncertainties are large. The model of \citeA{noll18b} also predicts a 
flattening of the height increase for high $N^{\prime}$. Moreover, this
model gives nearly constant heights for the lowest $N^{\prime}$. The latter 
is something which we do not find in our analysis that suggests a linear 
increase due to the growing contribution of the hot population. The nearly 
constant height for the OH lines with the highest $\Delta E^{\prime}$ of all 
$v^{\prime}$ also seems to be supported by the measured wave amplitudes. The 
interesting levels are occupied by a pure hot population \cite{noll20}, which
appears to show a nearly constant amplitude for the Q2DW in 2017 
(Figure~\ref{fig:xs_ampl_phase}). Moreover, the SABER-based fits of the 
profiles indicate a steep gradient of the amplitude at the heights relevant 
for OH (section~\ref{sec:profiles}). Hence, an almost constant amplitude 
would require a relatively narrow altitude distribution for the OH lines 
dominated by the hot population. For lower $N^{\prime}$, the interpretation 
of the wave amplitudes is more difficult due to the mixing of cold and hot 
populations with different altitude distributions.

\subsection{Impact of Time Range}\label{sec:2019}

\begin{figure}
\includegraphics[width=20pc]{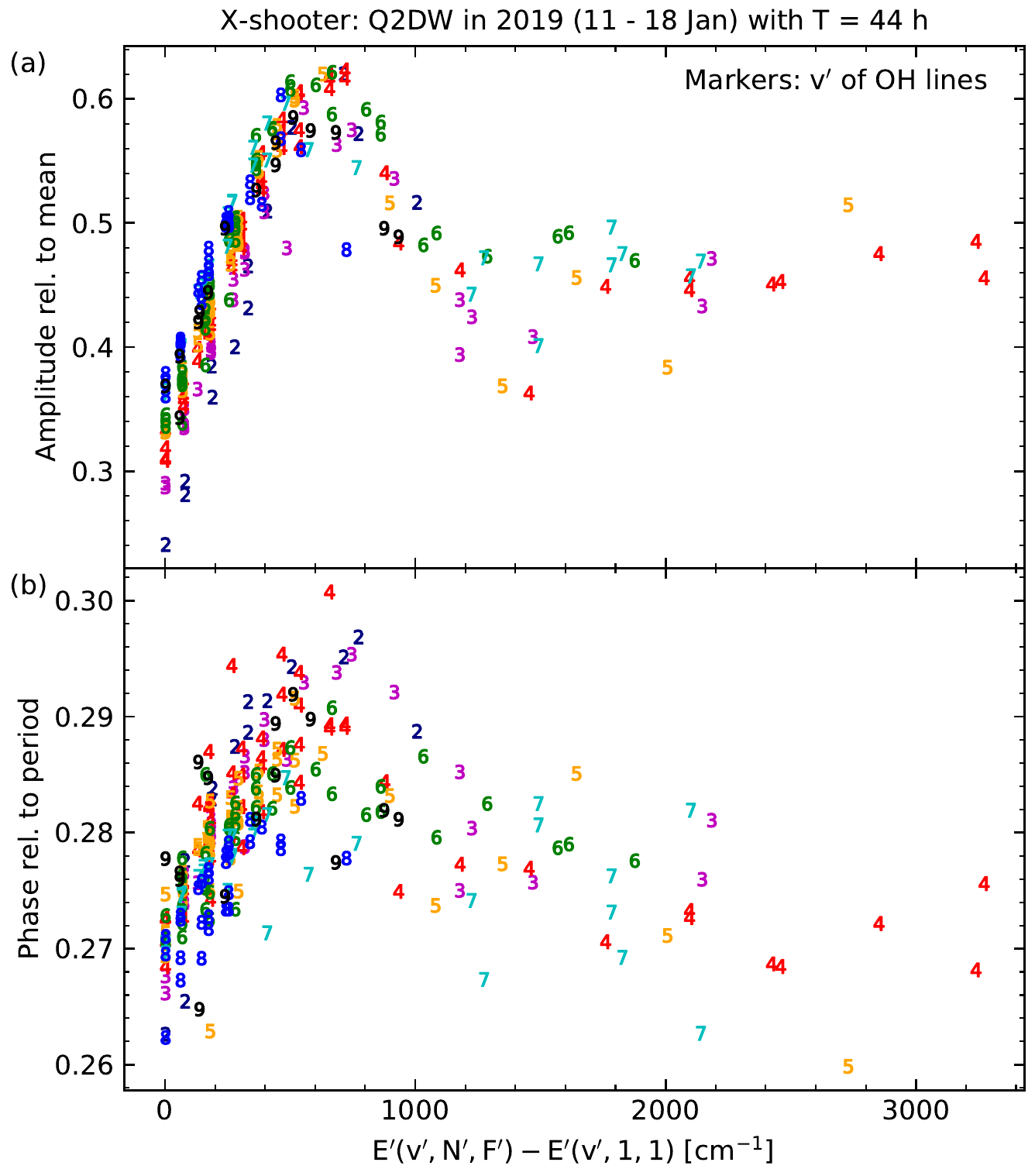}
\caption{Maximum amplitude $c{\cdot}a$ relative to the mean (a) and phase 
$\phi$ relative to the period at 15 January 12:00 LT at Cerro Paranal (b) for
all 270 OH lines used for the fit of the Q2DW event in 2019. The plot is 
similar to Figure~\ref{fig:xs_ampl_phase}.}
\label{fig:xs_ampl_phase_19}
\end{figure}

\begin{figure}
\includegraphics[width=20pc]{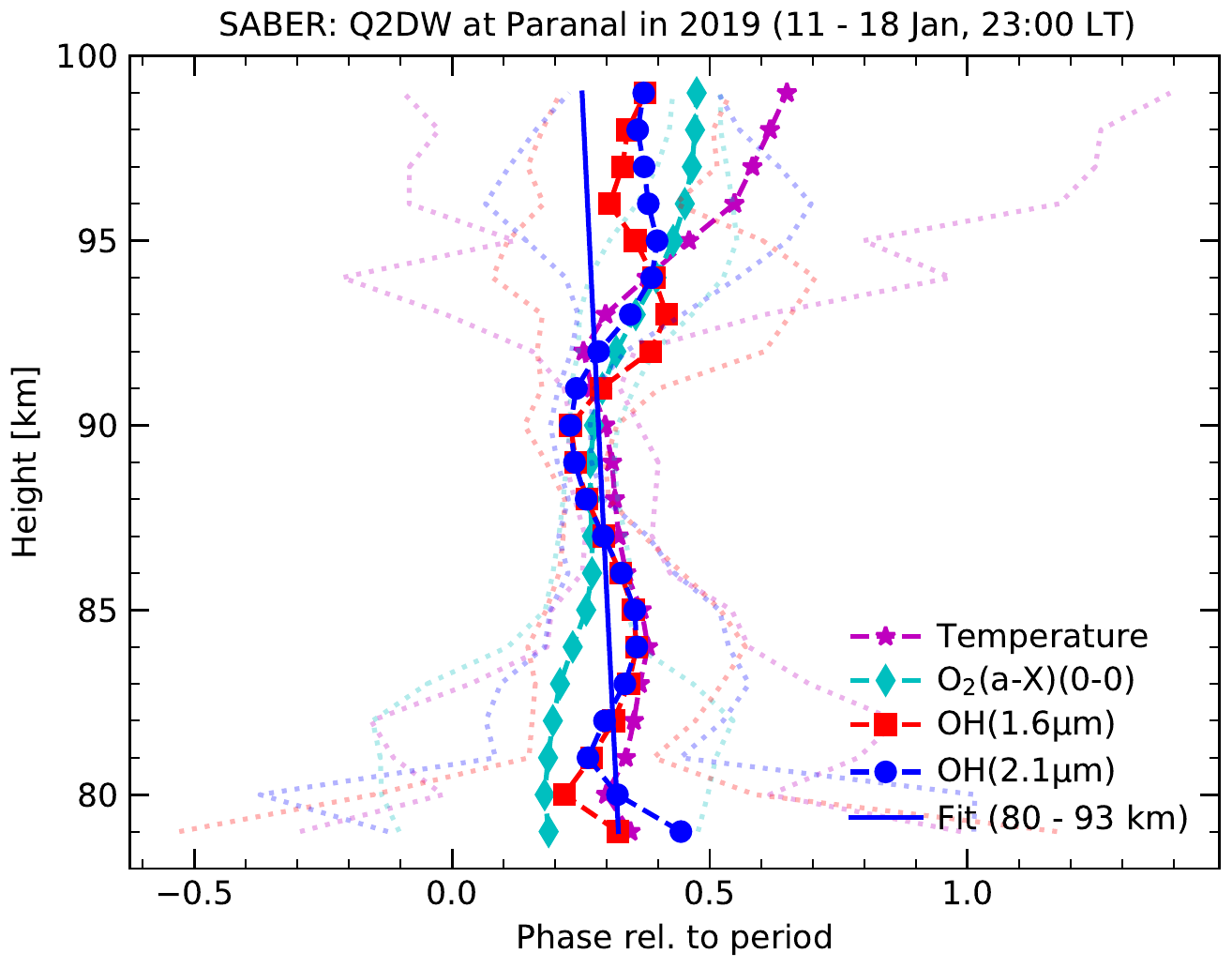}
\caption{Fitted phase relative to the period of 44\,h as a function of 
height for the SABER product kinetic temperature (stars) and the VERs of 
O$_2$(a-X)(0-0) at 1.27\,$\mu$m (diamonds), OH(1.6\,$\mu$m) (squares), and 
OH(2.1\,$\mu$m) (circles). The time series comprised the 19 limb scans taken
at about 23:00 LT in seven nights of 2019. The reference time for the plotted
phases was 15 January 12:00 LT at Cerro Paranal. Fit uncertainties, which are
useful for relative quality comparisons, are indicated by dotted lines. 
Moreover, the plot shows the resulting regression line for a linear fit of 
the OH(2.1\,$\mu$m) phases in the altitude range between 80 and 93\,km.}
\label{fig:sb_phases_19}
\end{figure}

Our OH height estimates are based on eight nights of a single Q2DW. Although
they appear to be reliable according to the discussion in 
section~\ref{sec:heights}, the analysis of another wave event would be a good
quality check. As already described in section~\ref{sec:xshooter}, we were 
able to identify another Q2DW in the X-shooter data of seven nights from 11 
to 18 January 2019. We analyzed the corresponding line measurements in the
same way as for the Q2DW in 2017. However, we only considered lines with
wavelengths shorter than 2.1\,$\mu$m because of too few spectra taken without
a $K$-blocking filter (section~\ref{sec:xshooter}). The resulting fits showed
that a period of 44\,h also appears to be the best choice. On the other hand,
the LT dependence of the wave amplitude was very different from the curves in
Figure~\ref{fig:xs_ampl_vs_hour}. The variation was much smaller. The maximum
and minimum $c{\cdot}a$ values only differed by a factor of 2 for the two 
example lines. Moreover, the highest amplitudes were reached between 22:00 
and midnight, which corresponds to a shift of $-2$\,h compared to the data 
from 2017. While the maximum relative amplitude for OH(4-2)P$_1$(14) did not 
change much (0.49 vs. 0.46 for 2017), it was significantly lower for 
OH(4-2)P$_1$(1) (0.31 vs. 0.74 for 2017). As a consequence, hot populations 
obviously showed a stronger response to the Q2DW than cold populations, which
is the opposite situation compared to 2017. This reversal was also seen for 
the dependence of the amplitude on $v^{\prime}$ for low $N^{\prime}$ 
(Figure~\ref{fig:xs_ampl_phase_19}a). Nevertheless, the impact of the 
mixing of both populations at intermediate rotational energies (400 to 
800\,cm$^{-1}$) was similar. The related lines showed the largest amplitudes,
although only values up to about 0.62 were found. In conclusion, the 
two-population model appears to be confirmed by the data from 2019. However, 
the Q2DW was weaker and showed different amplitude relations, which suggests 
that the properties of such waves are highly variable due to changes in their
generation, propagation, and interaction with the background atmosphere 
(including other waves). This high amount of variability had already been 
observed before \cite<e.g.,>{ern13,gu19,tunbridge11}.

For our purpose, it is important that a wave can be used for estimates of
effective emission heights. Hence, the phase relations are crucial. 
Unfortunately, the pattern for our set of OH lines was completely different
from the situation in 2017 (Figure~\ref{fig:xs_ampl_phase}b). A clear
$v^{\prime}$ dependence was not found and there was no monotonic 
decrease of the phase with increasing $v^{\prime}$ 
(Figure~\ref{fig:xs_ampl_phase_19}b). Instead, the phase even increased 
up to about 600\,cm$^{-1}$. The expected behavior was only present for higher 
energies. An important detail for the explanation of this unexpected 
structure is the maximum range of phases, which amounts to only 0.041. The 
standard deviation was 0.007. Consequently, the phase was almost constant. We 
also fitted the available SABER data for a better understanding. For the 
seven nights in 2019, we could use 19 profiles, all taken at about 23:00 LT 
(section~\ref{sec:saber}). The profile fits for both OH channels agree quite 
well with the X-shooter data. There were only small phase changes without 
clear direction (Figure~\ref{fig:sb_phases_19}). For OH(2.1\,$\mu$m), we 
found a maximum phase difference of only 0.13 for the height interval between 
80 and 93\,km, which minimizes the deviation of the fitted phases from a 
linear relation. The resulting regression line is almost vertical with a 
highly uncertain wavelength $\lambda_z$ of $280 \pm 240$\,km. The rising wave
turns into a descending one with $\lambda_z = 300 \pm 210$\,km for a fit up 
to 97\,km as in the case of 2017 (Figure~\ref{fig:sb_phases}). Hence, the
propagation direction remains unclear. The long wavelength may explain the 
reduced dependence of the wave amplitude on the OH line parameters (partly 
related to the emission height) and local time. The latter might point to a 
less efficient interaction of the Q2DW with the migrating diurnal tide, which 
has a relatively short $\lambda_z$ \cite<e.g.,>{forbes95} that is similar to 
the about 32\,km for our best fit of the Q2DW in 2017.

\citeA{huang13} investigated $\lambda_z$ of southern W3 Q2DWs based on SABER
temperature data from 2002 to 2011 and found a wide range of possible values 
from above 10\,km to beyond 100\,km at an altitude of 85\,km. As other 
studies using different techniques also found a high variability for 
low-to-middle southern latitudes in the mesopause region 
\cite{ern13,guharay13,reisin21}, the differences in $\lambda_z$ for the 
analyzed Q2DWs in 2017 and 2019 do not appear to be uncommon. For a strong 
event in southern summer 2002 to 2003, \citeA{huang13} also investigated the 
change of the wavelength over the lifetime of the wave of several weeks. They
found a trend of decreasing $\lambda_z$ and less variability with increasing 
age. As we investigated data from 11 to 18 January in 2019 but from 26 
January to 3 February in 2017, this trend would be consistent with our wave 
fits. However, as the Q2DWs can be quite different from year to year, a 
convincing check would need a detailed wave analysis over the entire lifetime
for the different years. In any case, the properties of the Q2DW in the 
selected time interval in 2019 were not suitable for our phase-sensitive 
investigation.

\section{Conclusions}\label{sec:conclusions}

In our study, we derived reference centroid emission heights for average
conditions of various individual OH lines for the first time. This success 
required the combination of OH line intensities from ground-based spectra 
taken with the astronomical X-shooter spectrograph at Cerro Paranal in Chile 
and space-based limb sounding of emission profiles in the two OH-related 
channels of the SABER radiometer on TIMED. Moreover, we benefited from the 
observation of a strong quasi-2-day wave (Q2DW) with both instruments in 
eight nights at the beginning of 2017. For the region around Cerro Paranal 
and the given observing period (mainly limited by the X-shooter data 
coverage), our wave fits of both data sets (separated depending on local
time) with a cosine function revealed a most likely period of 44\,h and 
vertical wavelength of about 32\,km (based on the SABER channel centered on 
2.1\,$\mu$m), which makes this wave event very suitable for phase-sensitive 
investigations. The amplitudes strongly varied. Our fits revealed 
particularly high amplitudes up to almost 100\% of the mean intensity for 
emission lines related to intermediate rotational energy between 400 and 
800\,cm$^{-1}$ and low vibrational upper level $v^{\prime}$, the second half 
of the night, and altitudes below the emission peak. The high values for
lines with intermediate rotational energies indicate an amplification of the 
variation due to the mixing of cold (thermalized) and hot (non-thermalized) 
OH rotational level populations, which maximizes for the stated energy range.
The local time dependence (with no wave detection at the beginning of the 
night) suggests that the Q2DW was strongly affected by the changing diurnal 
atmospheric conditions (e.g. by tides). Apart from the altitude dependence of
the intrinsic amplitude of the upward-propagating wave, the OH emission 
should also be affected by the increasing relative atomic oxygen variability 
with decreasing height. Furthermore, the wave properties significantly depend
on the selection of the geographical area and the time range. Another Q2DW 
which was present in the X-shooter data of 2019 indicated a vertical 
wavelength being too long to provide sufficient phase sensitivity for our 
purpose.

From the effective wave phases of each line measured by X-shooter and the 
relation between phase and altitude from the height-resolved fits of the 
SABER profiles, we first estimated effective emission heights that are 
representative of the altitudes with the strongest wave amplitudes during
the studied eight nights in 2017. For OH(2.1\,$\mu$m), the phase change 
between 80 and 97\,km was almost perfectly linear, which allowed us to derive
reliable heights without ambiguities. The resulting heights of the 298 
investigated OH emissions cover a range of about 8\,km with an average of 
84.9\,km. Lines with higher $v^{\prime}$ and/or rotational upper level 
$N^{\prime}$ show higher effective altitudes. At low rotational energies, the 
height increase appears to be almost linear, whereas lines with high 
$N^{\prime}$ indicate a flattening of the trend and a decreasing difference 
between different $v^{\prime}$. The latter could imply the presence of a 
universal hot population. Finally, we derived line-specific reference 
altitudes that are representative of the long-term centroid heights at Cerro 
Paranal. In combination with results for both SABER OH channels from a 
previous study, we found that a fixed positive shift of about 4.1\,km is 
obviously sufficient for our line set. The resulting heights therefore range 
from 85.9 to 93.8\,km with an average of 88.9\,km.       
   
Our results may provide important constraints for a better modeling of the
layering of OH emission. Moreover, the line-dependent heights could be used 
to study other wave events where suitable profile data are not available. In
particular, gravity waves with their too short periods for repeated 
satellite observations may constitute an appealing target. Finally, our 
approach could also be applied to other airglow emissions. The X-shooter 
spectra contain various candidates. Consequently, the results of our study
and possible future applications are quite promising with respect to a better
understanding of the chemistry and dynamics of the Earth's mesopause region.

\section*{Open Research}



The basic X-shooter data for this project originate from the ESO Science 
Archive Facility at \mbox{http://archive.eso.org} and are related to 
different observing programs. In particular, raw NIR-arm spectra taken 
between 26 January and 3 February 2017 and between 11 and 18 January 2019 
were processed (using the corresponding calibration data) and then analyzed. 
This project made also use of SABER v2.0 limb-sounding products at 
\mbox{http://saber.gats-inc.com} from January and February of the years 2017 
and 2019. Both archives can be accessed after registration.

The input data for the fitting procedure described in 
section~\ref{sec:methods} and the results as shown in the figures are
available at the public repository Zenodo \cite{noll22ds}. In detail, the data
release includes time series of the intensities of the investigated OH 
lines measured in the X-shooter spectra for the analyzed periods binned in 
30\,min steps. The corresponding SABER profiles for the two OH channels as 
well as the vertically integrated OH emissions are also provided as time 
series. Moreover, there are tables with the plot data of the 11 figures.
For the Q2DW from 2019, where only two figures are included in the paper, some
additional tables similar to the ones for the Q2DW from 2017 are considered.

\acknowledgments
Stefan Noll is financed by the project NO\,1328/1-3 of the German Research
Foundation (DFG). We thank Holger Winkler from Universit\"at Bremen for his 
contribution to the discussion and Sabine M\"ohler from ESO for her support 
with respect to the X-shooter calibration data. Moreover, we are grateful to
the three anonymous reviewers for their valuable comments.


%
%



\bibliography{Nolletal2022a}

%
%
%
%
%

\end{document}